\input harvmac
\input pictex
\def\neqno#1{\eqnn{#1} \eqno #1}
\def\M{{\cal M}}
\def\Mt{\tilde{\cal M}}
\def\Tr{\mathop{\rm Tr}}
\newbox\phdr
\setbox\phdr=\hbox{\beginpicture
\setcoordinatesystem units <1.0truein,1.0truein>
\setquadratic
\plot
0 0
0.0375 -0.03125
0 -0.0625
-0.0375 -0.09375
0 -0.125 /
\endpicture}
\def\photondr #1 #2 *#3 /{\multiput {\copy\phdr}  at
#1 #2 *#3 0 -0.125 /}
\newbox\hphdr
\setbox\hphdr=\hbox{\beginpicture
\setcoordinatesystem units <1.0truein,1.0truein>
\setquadratic
\plot
0 0
-0.03125 0.0375 
-0.0625 0
-0.09375 -0.0375 
-0.125 0 /
\endpicture}
\def\hphotondr #1 #2 *#3 /{\multiput {\copy\hphdr}  at
#1 #2 *#3 -0.125 0 /}
%
\overfullrule=0pt

%

\Title{hep-ph/9908330, \#HUTP-99/A043}
{\vbox{\centerline{The Phenomenology of a Top Quark Seesaw Model}
 }}
\smallskip
\centerline{Hael Collins$^\dagger$,   
            Aaron Grant$^\ddagger$ and 
            Howard Georgi$^*$ } 
\smallskip
\centerline{\it Harvard University}
\centerline{\it Cambridge, MA 02138, USA}
\smallskip
\centerline{$^\dagger${\tt hael@feynman.harvard.edu} }
\centerline{$^\ddagger${\tt grant@feynman.harvard.edu} }
\centerline{$^*${\tt georgi@physics.harvard.edu} }

\bigskip

\medskip

\centerline{ABSTRACT}
\smallskip

\noindent
The top quark seesaw mechanism offers a method for constructing a composite
Higgs field without the usual difficulties that accompany traditional
technicolor or topcolor theories.  The focus of this article is to study
the phenomenology of the new physics required by this mechanism.  After
establishing a set of criteria for a plausible top quark seesaw theory, we
develop two models, the first of which has a heavy weak singlet fermion
with hypercharge ${4\over 3}$ while the second has, in addition, a heavy
weak singlet hypercharge $-{2\over 3}$ fermion.  At low energies, these
theories contain one or two Higgs doublets respectively.  We then derive
the low energy effective Higgs potential in detail for the two-doublet
theory as well as study the likely experimental signatures for both
theories.  A strong constraint on the one-doublet model is the measured
value of the $\rho$ parameter which permits the new heavy fermion to have a
mass of about $5$--$7\, {\rm TeV}$, when the Higgs has a mass greater than
$300\, {\rm GeV}$.  In the two-doublet model, mixing of the new heavy
$Y=-{2\over 3}$ fermion and the $b$ quark affects the prediction for $R_b$.
In order to agree with the current limits on $R_b$, the mass of this
fermion should be at least $12\, {\rm TeV}$.  The mass of the heavy
$Y={4\over 3}$ fermion in the two-doublet model is not as sharply
constrained by experiments and can be as light as $2.5\, {\rm TeV}$.

\Date{August, 1999}

\newsec{Introduction}

In recent years the standard model has been subjected to extraordinarily
precise experimental tests 
\ref\ewwg{The LEP Collaborations ALEPH, DELPHI, L3, OPAL, the LEP
Electroweak Working Group and the SLD Heavy Flavour and Electroweak Groups,
``A Combination of Preliminary Electroweak Measurements and Constraints
on the Standard Model,'' CERN-EP/99-15.}.  
All the evidence to date suggests that the usual picture of fundamental
interactions, based on a spontaneously broken $SU(2)\times U(1)$ gauge
symmetry, is quantitatively correct.  However, the character of the
symmetry breaking sector of the theory is still largely mysterious.
Constraints derived from precision electroweak data suggest that the Higgs
boson may be light $\ewwg$, although this conclusion has been criticized on
various grounds
\ref\chanowitz{M.~S.~Chanowitz, ``Combining real and virtual Higgs boson 
mass constraints,'' Phys.\ Rev.\ Lett.\  {\bf 80} (1998) 2521; 
{\tt hep-ph/9710308}.  
R.~S.~Chivukula and N.~Evans, ``Triviality and the precision bound on 
the Higgs mass,'' {\tt hep-ph/9907414}.}.  
In addition, it has been argued 
\ref\peskin{M.~E.~Peskin and T.~Takeuchi, ``Estimation of oblique 
electroweak corrections,'' Phys.\ Rev.\ {\bf D46} (1992) 381.}  
that models of electroweak symmetry breaking that involve large numbers of
new strongly interacting $SU(2)$ doublet fermions are excluded by
experiment.  These constraints pose a significant challenge to traditional
technicolor models 
\ref\technicolor{For a review, see E.~Farhi and L.~Susskind,
``Technicolor,'' Phys.\ Rept.\ {\bf 74} (1981) 277.}
since such models typically predict a heavy Higgs boson, and often involve
large numbers of new $SU(2)$ doublets.

These considerations suggest that the simplest technicolor mechanisms may
not be realized in nature.  An alternative scenario has been suggested 
\ref\bardeen{W.~A.~Bardeen, C.~T.~Hill and M.~Lindner, ``Minimal 
dynamical symmetry breaking of the standard model,'' Phys.\ Rev.\ {\bf D41}
(1990) 1647},
in which the composite Higgs field is ``made'' of ordinary standard model
quarks.  Indeed, the composite field $\bar{Q}_R \psi^{3}_L$, where
$\psi^{3}_L$ is the $(t,b)$ doublet and $Q$ is a quark with electric charge
$+2/3$, has the correct quantum numbers to play the role of the standard
model Higgs boson.  In models of this type, four-fermi operators that
result from integrating out physics at a high scales bind the composite
Higgs and break $SU(2)$ through the formation of a $\bar{Q}_R
\psi^{3}_L$ condensate.  The simplest models, in which $Q_R\equiv t_R$,
predict a top mass that is too large.  So it has been suggested that $Q_R$
is simply a new isosinglet quark with the quantum numbers of $t_R$, and
that the physical top quark mass comes about via a ``see-saw'' mechanism
involving $t$ and $Q$
\ref\dh{Bogdan A.~Dobrescu and Christopher T.~Hill, ``Electroweak Symmetry
Breaking via a Top Condensation Seesaw Mechanism,'' Phys.\ Rev.\ Lett.\
{\bf 81} (1998) 2634; {\tt hep-ph/9712319}.}.

A full description of the top-condensate see-saw mechanism must involve new
interactions at high scales.  In particular, the four-fermi operators
needed to bind the composite Higgs and break $SU(2)$ may come about as the
result of integrating out the massive gauge bosons of a spontaneously
broken ``topcolor'' gauge interaction 
\ref\topcolor{C.~T.~Hill, ``Topcolor: Top quark condensation in a gauge 
extension of the standard model,'' Phys.\ Lett.\ {\bf B266}, (1991) 419;
C.~T.~Hill, ``Topcolor assisted technicolor,'' Phys.\ Lett.\ {\bf B345} 
(1995) 483, {\tt hep-ph/9411426}.}.
So, in a sense, top-condensate models defer the problem of gauge symmetry
breaking to a new $SU(2)$ singlet sector of the theory, whose interactions
break topcolor.  

In the present article, we discuss some of the issues involved in
constructing topcolor models that are ``complete,'' in the sense that all
gauge symmetry breaking is accomplished dynamically, without recourse to
{\it ad hoc\/} phenomenological Higgs multiplets.  In Sec.~2, we review the
basic features of top condensate see-saw models.  In Sec.~3, we discuss
models of physics above the scale of topcolor breaking, and delineate two
classes of models that appear to be viable.  In Sec.~4 we discuss the
spectrum of composite Higgs bosons in these models.  In Sec.~5, we study
the phenomenology of these models at low energies, and present experimental
constraints on the masses of new particles predicted by our models.
Finally, Sec.~6 concludes.

\newsec{The Development of the Top Quark Seesaw Mechanism.}

As experiments placed larger and larger bounds on the mass of the top
quark, it became tempting to speculate that the top quark plays a principal
role in the breaking of electroweak symmetry.  This observation led
Nambu and others
\ref\topcond{Y.~Nambu, report EFI 88-39 (July 1988), published in the 
Proceedings of the {\it Kazimierz 1988 Conference on New theories in
physics\/}, ed.\ T.~Eguchi and K.~Nishijima; in the Proceedings of the {\it
1988 International Workshop on New Trends in Strong Coupling Gauge
Theories\/}, Nagoya, Japan, ed. Bando, Muta and Yamawaki (World Scientific,
1989); report EFI-89-08 (1989); V.~A.~Miransky, M.~Tanabashi and
K.~Yamawaki, Mod.\ Phys.\ Lett.\ {\bf A4} (1989) 1043; Phys.~Lett.~{\bf B
221} (1989) 177; W.~J.~Marciano, Phys.\ Rev.\ Lett.\ {\bf 62} (1989)
2793.}  
to suggest that if some new interactions produced a top quark condensate, 
$\langle\bar tt\rangle$, this condensate would have the correct gauge
properties to break $SU(2)_W \times U(1)_Y \to U(1)_{\rm em}$.  The
interaction that he studied was the four-fermion interaction,
$${\cal L} = {\cal L}_{\rm kinetic} 
+ G (\bar\psi_{La}t_R)(\bar t_R \psi_L^a). \neqno\njl$$
Here, $\psi_L$ is the usual left-handed third generation quark doublet and
$t_R$ is the right-handed top quark.  Bardeen, Hill and Lindner $\bardeen$,
following Nambu, examined this interaction further by performing a
renormalization group analysis of the low energy theory.  By summing the
graphs contributing to the $W$-propagator in the bubble approximation, they
estimated that the scale associated with electroweak symmetry breaking,
$v=246\, {\rm GeV}$, and the top quark mass are related by the
Pagels-Stokar formula
\ref\pagelsstokar{H.~Pagels and S.~Stokar, ``The Pion Decay Constant, 
Electromagnetic Form-Factor And Quark Electromagnetic Selfenergy In QCD,''
Phys.\ Rev.\ {\bf D20} (1979) 2947.}
$$v^2 \approx {N_c\over 8\pi^2} m^2_t \ln{\Lambda^2\over m_t^2}. 
\neqno\pagels$$
For $\Lambda$ of order 1 TeV, this indicates a top mass of order 600 GeV.
The prediction of $m_t$ can be lowered by taking the cutoff $\Lambda$ to be
very large.  However, even for $\Lambda \sim M_{\rm Planck}$, the predicted
top quark mass is too large.  Indeed, the detailed analysis of $\bardeen$
indicates that $m_t=218$ GeV for this value of $\Lambda$.

Apart from the top mass prediction, this theory for top condensation has
several unappealing features.  The interaction in $\njl$ is not
renormalizable and must be the low energy remnant of some new physics.
More seriously, the strength of the interaction must be very precisely
tuned in order to obtain $M_{W,Z} \ll \Lambda$. This is the usual gauge
hierarchy problem.

The first difficulty was resolved when Hill $\topcolor$\  noticed that,
through a Fierz rearrangement, the four-fermi interaction could be
rewritten in a suggestive form.  Indeed, we can make the substitution
$$
(\bar\psi_{La}t_R)(\bar t_R \psi_L^a) \to - (\bar\psi_{La}\gamma_\mu
{\textstyle{1\over 2}} \lambda^A \psi_L^a) (\bar t_R \gamma^\mu
{\textstyle{1\over 2}} \lambda^A t_R) + {\cal O}(N^{-1}_c). \neqno\fierz
$$
The operator on the right-hand side is precisely what one obtains from
integrating out a massive gauge boson.  If quantum chromodynamics is
embedded in a larger gauge theory, called ``topcolor'', then an interaction
such as that on the right side of $\fierz$ naturally arises.  For example,
suppose that the topcolor gauge symmetry is $SU(3)_1\times SU(3)_2$, which
spontaneously breaks to the diagonal subgroup. The theory contains eight
massless gauge fields, the gluons, and eight gauge fields of mass $M$, the
colorons.  With appropriate gauge quantum number assignments, the exchange
of a massive coloron produces the operator on the right side of $\fierz$
when we integrate it out.  If $g_{\rm tc}$ is the gauge coupling constant,
we identify $G\sim g^2_{\rm tc}/M^2$.

The Pagels-Stokar relation suggests that if the scale $\Lambda$ of new
physics is in the TeV range, the top quark mass resulting from $SU(2)$
breaking will be unacceptably large.  A solution to this difficulty $\dh$\  
is to introduce a new heavy fermion, $\chi$, a weak singlet with the same
hypercharge as $t_R$, which participates in the breaking of electroweak
symmetry.  Since $\chi$ is an isosinglet, it does not contribute too much
to Peskin and Takeuchi's $S$ parameter $\peskin$.  By comparison, in
traditional technicolor models, the large number of ``extra'' $SU(2)$
doublets needed to construct the Higgs sector give large contributions to
$S$ that are probably unacceptable.

The introduction of the $\chi$ leads to a modification of the Pagels-Stokar
relation for the top quark mass.  We can derive the new relation using an
effective field theory approach
\ref\seesaw{R.~Sekhar Chivukula, Bogdan A.~Dobrescu, Howard
Georgi, Christopher T.~Hill, ``Top Quark Seesaw Theory of Electroweak
Symmetry Breaking,'' Phys.\ Rev.\ {\bf D59} (1999) 075003; {\tt
hep-ph/9809470}.}.  
To begin, we assume that the gauge structure of the theory is such that the
exchange of a massive coloron between $\psi_L$ and $\chi_R$ (rather than
the $t_R$) produces a four-fermion interaction of the form
$$
{\cal L}' = {g^2_{\rm tc}\over M^2}(\bar\psi_L\chi_R)(\bar\chi_R\psi_L)
+ \cdots , \neqno\fourf
$$
In addition, the theory admits, after topcolor symmetry breaking, the
following allowed mass terms
$$
{\cal L}' = - \mu_{\chi\chi} \bar\chi_L\chi_R 
- \mu_{\chi t}\bar\chi_L t_R + \hbox{h.c.} + \cdots . \neqno\masses
$$
Since both $\chi_L$ and $t_R$ are $Y={4\over 3}$ singlets, these fields can
mix, so we define mass eigenstates through the following rotation:
$$
\eqalign{
\chi'_R &= \cos\theta\, \chi_R + \sin\theta\, t_R \cr
t'_R &= \cos\theta\, t_R - \sin\theta\, \chi_R \cr}
\qquad\qquad
\tan\theta = {\mu_{\chi t}\over \mu_{\chi\chi} } .
\neqno\mrotate$$
The interaction Lagrangian then becomes
$$
{\cal L}' = - \overline{m} \bar\chi_L\chi'_R + \hbox{h.c.}
+ {g^2_{\rm tc}\over M^2} 
\bigl( \bar\psi_L (\cos\theta\, \chi'_R - \sin\theta\, t'_R) \bigr)
\bigl( (\cos\theta\, \bar\chi'_R - \sin\theta\, \bar t'_R)\psi_L \bigr) 
\neqno\rotlagr
$$
where ${\overline{m}}^2 \equiv \mu^2_{\chi\chi} + \mu^2_{\chi t}$.  We can
analyze the effects of the four-fermi interaction by rewriting the
Lagrangian in terms of a static auxiliary Higgs field $H$.  The
interaction Lagrangian can be written as
$$
{\cal L}' = - \overline{m} \bar\chi_L\chi'_R + \hbox{h.c.} 
+ g_{\rm tc} \bar\psi_L (\cos\theta\, \chi'_R - \sin\theta\, t'_R)H_0
+ \hbox{h.c.} - M^2 H_0^\dagger H_0 ; \neqno\auxh
$$
at low energies, we shall find that $H_0$ plays the role of the
unrenormalized Higgs doublet.  For energies $\mu<\overline{m}<M$, the
$\chi$ field decouples and, upon integrating it out of the theory,
generates the following one loop effective Lagrangian:
$$
{\cal L}'_{\rm eff} 
= - g_{\rm t}\sin\theta \left[ \bar\psi_L t'_R H + \hbox{h.c.} \right] 
+ D_\mu H^\dagger D^\mu H - m_H^2 H^\dagger H 
- \lambda (H^\dagger H)^2 . \neqno\prophiggs
$$
Here, $H$ is the renormalized Higgs field, $H= \sqrt{Z_H}H_0$, and $g_{\rm
t}$ is the renormalized coupling, $g_{\rm t} = g_{\rm tc}/\sqrt{Z_H}$,
where the wave function renormalization induced by integrating out the
$\chi$ is
$$
Z_H = {g^2_{\rm tc} N_c\over 16\pi^2} \left[ \ln{M^2\over\overline{m}^2}
+ \sin^2\theta \ln{\overline{m}^2\over \mu^2} + {\cal O}(1) \right] .
\neqno\wfrenorm
$$
The Higgs mass and self-coupling constant, in terms of the unrenormalized
quantities $m_{H_0}^2=Z_H m_H^2$ and $\lambda_0=Z_H^2\lambda$, are
$$
\eqalign{
m^2_{H_0} &= M^2 - {g_{\rm tc}^2 N_c\over 8\pi^2} \left[ M^2 - \cos^2\theta
\overline{m}^2 \ln{M^2\over\overline{m}^2} \right] 
+ {\cal O}(\overline{m}^2,\mu^2) \cr
\lambda_0 &= {g_{\rm tc}^2 N_c\over 8\pi^2} \left[ \ln{M^2\over\overline{m}^2} 
- \sin^4\theta\ln{\overline{m}^2\over\mu^2} + {\cal O}(1) \right] .\cr}
\neqno\mlrenorm$$
Observe that for $\mu<\overline{m}<M$, $H$ is a dynamic scalar field which
has the same quantum numbers as a Higgs field.  If this dynamically
generated mass term becomes negative, then the Higgs acquires a vacuum
expectation value.  This occurs for $\cos\theta\approx 1$ when
$${g^2_{\rm tc}N_c\over 8\pi^2} \ge
\left[ 1 - {\mu_{\chi\chi}^2\over M^2} 
\ln{M^2\over\mu_{\chi\chi}^2} \right]^{-1} $$
where we have used 
$$\overline{m}^2 = \mu^2_{\chi\chi} + \mu^2_{\chi t} \approx
\mu^2_{\chi\chi}$$
which is appropriate for $\sin\theta \ll 1$.  If we then write the Higgs 
vacuum expectation value in the usual form,
$$\langle H\rangle = \pmatrix{v/\sqrt{2}\cr 0\cr} \qquad v=246\, {\rm GeV},
\neqno\higgsvev$$
the top quark acquires a mass,
$$m_t = g_{\rm t} \sin\theta {v\over\sqrt{2}}. \neqno\mt$$
The coupling $g_{\rm t}$ is related to the topcolor gauge coupling by a
factor of the wave function renormalization: $g_{\rm t} = g_{\rm tc}/
\sqrt{Z_H}$.  When we substitute in the expression in equation
$\wfrenorm$ and retain only the leading term in $\sin^2\theta$, we have an
expression which superficially resembles that of the bubble approximation
$\pagels$:  
$$v^2 = {N_c\over 8\pi^2} {m^2_t\over\sin^2\theta} 
\ln{M^2\over\overline{m}^2} + {\cal O}(\sin^2\theta). \neqno\betterps$$
Therefore we see that it is possible naturally to have a top quark
much lighter than the $600\, {\rm GeV}$ required by the bubble
approximation formula:
$$m_t = 174\, {\rm GeV} \qquad\quad
{m_t\over\sin\theta} \sim 600\, {\rm GeV} . \neqno\massseesaw$$

Up until now, we have made several assumptions about the relative sizes of
the mass terms as well as which couplings appear in the low energy
effective Lagrangian.  For these to occur naturally constrains the models
that we can consider.  For example, equation $\betterps$ requires that
$\sin\theta\ll 1$, or $\mu_{\chi\chi}\gg \mu_{\chi t}$, so the dynamics
that produce these mass terms should naturally favor a heavier mass for the
$\bar\chi\chi$-term.  Moreover, to produce a successful seesaw, we
neglected terms containing $\bar\psi_L t_R$, so such terms should also be
naturally suppressed.  For a specific model, these restrictions on the
sizes of the mass terms translate into the requirements on the relative
mass dimensions of the effective operators that produce them in the higher
energy theory.

By a completely analogous procedure, we could also introduce a weak singlet
partner for the $b$-quark,
$$\omega\sim \left( 1, - {\textstyle{2\over 3}} \right)\quad \hbox{under}
\quad SU(2)_W \times U(1)_Y. \neqno\introw $$
The low energy spectrum of such a theory, which we study in section 4,
contains two Higgs doublets.  In models of this type, the top and bottom
quark masses can be generated by the $SU(2)$ breaking condensates $\langle
\bar{t}_L \chi_R \rangle$ and $\langle \bar{b}_L \omega_R\rangle$.  Of 
course, it is necessary to adjust the see-saw mechanism appropriately in
order to get the correct $b$ quark mass.  This is different from the
situation in models with only a $\chi$-type quark.  In models with only a
$\chi$, it is often necessary to ``tilt'' the vacuum in such a way that
only $t$-quark condensates form; if the $b$ condenses as well, its mass
becomes too large.  From this perspective, models with both $\chi$ and
$\omega$ quarks have the virtue that they do not require such tilting
mechanisms.

\newsec{Constructing a Successful Topcolor Seesaw Model.}

The gauge theory structure of a successful topcolor theory is of the
general form
\eqn\gauge{ G\times G_{\rm tc}\times SU(2)_W \times U(1)_Y }
where $G_{\rm tc}$ is the topcolor group, usually two or more copies of
$SU(3)$, that breaks down to ordinary $SU(3)_{\rm color}$ under the
influence of some additional gauge interactions with the local symmetry
group $G$.  In the simplest topcolor models, $G_{\rm tc} = SU(3)\times
SU(3)$, but we shall later study a model with three $SU(3)$ factors.  The
matter content of the theory should include, in addition to the standard
model fields, some $SU(2)_W$-inert fermions $\chi_{L,R}$ with hypercharge
$Y={4\over 3}$ (and perhaps some $Y=-{2\over 3}$ weak singlets for a $b$
quark seesaw), and some fermions that break topcolor when the $G$ gauge
interactions become strong.  Some additional matter fields may be required
to cancel the anomalies in the theory.  A realistic topcolor seesaw model
must be arranged to satisfy the following properties to produce the correct
low energy physics.  We shall assume that the models are self-contained to
the extent that they are anomaly free and do not require unspecified
``spectator'' fermions to cancel gauge anomalies.  In addition, in a fully
realistic model, it must be possible to construct higher-dimensional
gauge-invariant operators that give rise to light quark and lepton masses.
We assume that such operators come from integrating out new physics at a
``flavor'' scale $\Lambda_f$ of order 50 to 100 TeV.  We shall further
assume that the flavor dynamics are strong.  This assumption is not
necessary but it is convenient since it allows us to use the tools of
na\"\i ve dimensional analysis (NDA)
\ref\nda{H.~Georgi, ``Generalized Dimensional Analysis,'' Phys.\ Lett.\
{\bf B298} (1993) 187; {\tt hep-ph/9207278}.}  
to estimate the mass scales of the effective operators that arise at
energies below the scales associated with the flavor physics, $f_f$ and
$\Lambda_f$.  Here $f_f$ is the decay constant of the pions associated with
flavor symmetry breaking, while $\Lambda_f \leq 4 \pi f_f$ is the physical
mass of the light (non-Goldstone) composite states associated with the
strong flavor interactions.

The requirement that it be possible for such operators to generate quark
and lepton masses of order 1 GeV can be used as a guide in building models.
We also demand that all the energy scales and particle masses should arise
dynamically.  This condition produces a natural hierarchy of particle
masses.  If the masses are not protected by chiral symmetries, it is
difficult to explain why they should be small compared to $\Lambda_f$.

Before stating the models we study here, we shall present several simpler
models since it is instructive to see how these fail.  The models are
represented in ``Moose notation'' 
\ref\moose{H.~Georgi, ``A Tool Kit For Builders Of Composite Models,''
Nucl.\ Phys.\ {\bf B266} (1986) 274; 
H.~Georgi, ``$SU(2)\times U(1)$ breaking, compositeness
and flavor,'' Les Houches, 1985 (North-Holland, 1987) 339.},
which efficiently encodes the gauge transformation properties of the matter
fields and allows us quickly to write anomaly-free models.  In the Moose
notation, the $SU(N)$ gauge groups are represented by circles while
fermions are lines.  A fermion line emerging from an $SU(N)$ group lives in
the $N$-representation if left-handed ($\overline{N}$ if right-handed)
while an entering fermion line lives in the $\overline{N}$-representation
if left-handed ($N$ if right-handed).

One might hope to succeed with a seesaw model involving a single group
mediating between the two $SU(3)$'s that compose the topcolor group.  An
example of such a theory is shown in figure 1.  In this model, the $SU(4)$
group can break the $SU(3)\times SU(3)$ symmetry dynamically.  In
particular, if the $SU(4)$ is more strongly coupled than the $SU(3)$
interactions, we expect the $\xi$ quarks to condense in the $\bar{\xi}_L
\xi_R$ channel.  This condensate transforms as $(3,\bar{3})$ under
$SU(3)\times SU(3)$ and therefore can break the $SU(3)$ factors down to
their diagonal $SU(3)$ subgroup.  Notice that the light fermions---the
$\chi$, $\omega$, and standard model fields---are anomalous under
$SU(3)\times SU(3)$.  By choosing the group $SU(4)$ for the interactions
that break topcolor, the $\xi_{L,R}$ fields cancel the topcolor anomalies.
To protect the light quark masses, we have chosen the left- and right-hand
fields to transform under different $SU(3)$ groups.  In this model, the
third generation is distinguished as follows:  we can choose $t_R$ and $b_R$
to be respectively the linear combinations of $U_R^i$ and $D_R^i$ quarks
that couple to the $\chi_L$ and $\omega_L$ fermions.  The left-handed
$\psi_L^3$ fields are defined by the linear combination of the $\psi_L^i$
fields that couples to the Higgs field that develops a vacuum expectation
value.  It is not clear, however, that prohibitively large flavor changing
neutral currents do not arise in this model.

The light quark masses arise from operators such as 
\eqn\seesawb{ {1\over \Lambda_f f_f^4} (\bar\psi_L^3\chi_R)
(\bar U_R^1\xi_L)(\bar\xi_R\psi_L^1). }
When the $SU(4)$ interactions cause the $\bar\xi_R\xi_L$ condensate to form
and break $SU(3)\times SU(3)\to SU(3)_{\rm QCD}$, the na\"\i ve estimate
for the a light quark mass is
\eqn\seesawc{ m_u \sim {\langle\bar\xi_R\xi_L\rangle\over \Lambda_f f_f^4} 
\langle\bar\psi^3_L\chi_R\rangle. }
$\langle\bar\xi_R\xi_L\rangle$ is the vacuum expectation value of the
$\bar\xi_R\xi_L$ condensate which is of the order
$\langle\bar\xi_R\xi_L\rangle \sim f_{\rm tc}^2 \Lambda_{\rm tc}$, using
the rules of na\"\i ve dimensional analysis $\nda$.  Here $f_{\rm tc}$ and
$\Lambda_{\rm tc}$ play the analogous roles for the strongly interacting
topcolor dynamics as $f_f$ and $\Lambda_f$ played in the flavor physics.
In terms of the coloron mass $M$ introduced in section 2, we have $M\sim
\Lambda_{\rm tc} \leq 4\pi f_{\rm tc}$.  One of the difficulties with this
model is that lepton masses can arise from dimension six operators,
\eqn\seesawd{ \epsilon_{ab}(\bar\psi_L^a\chi_R)(\bar\ell_L^b e_R) }
and should be generically heavier than the quark masses by a factor
$\Lambda_f f_f^2/\langle\bar\xi_R\xi_L\rangle$.  For example, when the
scale of the flavor physics is $f_f\sim 100\, {\rm TeV}$ and $f_{\rm
tc}\sim 10\, {\rm TeV}$, this factor is ${\cal O}(10^3)$ which is
unacceptably large.  Another difficulty for models with $G_{\rm
tc}=SU(3)\times SU(3)$ is that no symmetry prevents a $\bar\chi_L t_R$ (or
even a $\bar\chi_L U_R^i$) mass term from arising.  Such a term could spoil
the seesaw mechanism since there is no reason that it could not have a mass
of the order of the flavor scale physics.  This observation suggests that
the fields $\chi_L$, $t_R$ and $\chi_R$ should transform under different
$SU(3)$ groups.  Here, we shall attempt to construct models which do not
admit these tree-level mass terms---a condition which will lead us to
consider models with more complicated gauge symmetries.  Yet the simplicity
of this linear Moose model is so enticing that we shall explore models
similar to it elsewhere
\ref\inprogress{H.~Georgi, A.~Grant and H.~Collins, hep-ph/9907477.}.

A model with three $SU(3)$ topcolor factors contains enough symmetry to
prevent the $\bar\chi_L t_R$ or $\bar\chi_R\chi_L$ terms from forming at
too high an energy scale.  One such model of this form is shown in figure 2 
where the dimensions of the $SU(m)$ groups that break topcolor have been
chosen to cancel any anomalies.  Unfortunately, in this model the 
$\bar\chi_L t_R$ mass term originates from a dimension nine operator:
\eqn\seesawf{ {\Lambda_f \over f_f^6} (\bar\chi_L\xi_R)
(\bar\xi_L\zeta_R)(\bar\zeta_L t_R)
\to {\Lambda_f\langle\bar\xi_L\xi_R\rangle \langle\bar\zeta_L\zeta_R\rangle
\over f_f^6}\, \bar\chi_L t_R 
\sim {\Lambda_f \Lambda^2_{\rm tc} f^4_{\rm tc}
\over f_f^6}\, \bar\chi_L t_R  
\qquad\hbox{(NDA)} }
which is probably too small unless the ratio of the flavor scale to the
topcolor scale is only about a factor of three.  This difficulty comes from
the need to straddle the entire diagram to produce a operator containing
$\chi_L$ and $t_R$ that is invariant under all of the gauge symmetries.
The remedy is to add another gauge group which links the two ends.  Thus we
are led to consider topcolor models such as that of figure 3 which
was first proposed in $\seesaw$.

With only a single additional ($Y={4\over 3}$) fermion, this model produces
a single Higgs $SU(2)_W$ doublet in the low energy theory.  An unpleasant
feature of this model is that it still requires some mechanism to tilt the
vacuum to prevent the formation of a large $\bar b_R \psi_L^3$ condensate
which would produce an unacceptably large $b$ quark mass \seesaw.  Figure 4
shows a model with two additional fermions, $\chi$ ($Y={4\over 3}$) and
$\omega$ ($Y=-{2\over 3}$), which act like a weak-inert fourth generation
and avoid this need for tilting.  At low energies, this model contains two
Higgs doublets corresponding to the $\bar\psi_L^3\chi_R$ and
$\bar\psi_L^3\omega_R$ condensates.  The $b$ quark mass participates in its
own seesaw so that it is possible to have $m_\omega\sim 10\, {\rm TeV}$
with $m_b \sim 4\, {\rm GeV}$.

To summarize, the desire to achieve an anomaly-free model that yields a
realistic low energy theory though operators of the appropriate mass
dimension has led us to consider models with a rather rich gauge group
structure.  We shall focus in particular upon the two models shown in
figures 3 and 4 but any topcolor model with a single $Y={4\over 3}$ fermion
or a $Y=\left( {4\over 3},-{2\over 3}\right)$ pair should share the same
general behavior of two triangular models above, in particular the bounds
on the masses of these fermions set by $Z\to b\bar b$ and the $\rho$
parameter discussed below.  Models with more $SU(2)_W$ singlet fermions, to
a first approximation, lead to multiple copies of the Higgs fields of these
two models which generically should reinforce the perturbations to $R_b$
and $\delta\rho$.  Some of our assumptions may be relaxed to obtain simpler
models but only at the cost of the naturalness of the mass scales.

\newsec{The Higgs Potential.}

Now that we have a pair of models that produce a top quark seesaw, we
would like to study their phenomenology in some detail.  The effective
potential for the one-doublet model has been carefully studied before
\seesaw\ so in this section we concentrate on the two-doublet model and
derive its Higgs spectrum.  A few points deserve special attention.  First,
in the leading approximation, the two doublet model preserves a global
Peccei-Quinn $U(1)$ symmetry, so it has an unacceptable weak scale axion.
We shall add explicit, but small, Peccei-Quinn breaking terms to give this
``axion'' a mass.  Second, We shall make some simplifying assumptions about
the dependence of the low energy theory on $m_\chi$ and $m_\omega$.  When
$m_\chi=m_\omega$, the model has a custodial $SU(2)_R$ symmetry which is
reflected in the Higgs spectrum: both Higgs doublets acquire equal vacuum
expectation values, and the scalars are grouped into custodial $SU(2)$
multiplets.  We shall assume that this behavior persists for $m_\chi \neq
m_\omega$, provided that both are light compared to the topcolor scale.  We
will return elsewhere to the study of custodial $SU(2)$ breaking by
$m_\chi$ and $m_\omega$.

We shall use the techniques of $\bardeen,\dh$ to study the low energy Higgs
spectrum.  These methods are equivalent to the NJL approximation
\ref\njl{Y.~Nambu and G.~Jona-Lasinio, ``Dynamical Model Of Elementary 
Particles Based On An Analogy With Superconductivity. I,'' Phys.\  Rev.\ 
{\bf 122} (1961) 345.},  
which is sufficient for our purposes.

In the two doublet model, the leading-order four-fermion interaction
in $1/N_c$,
\eqn\higgsa{ {\cal L}_{\rm int}  = {g_{\rm tc}^2\over M^2} \left[
(\bar\psi_L\chi_R)(\bar\chi_R\psi_L) 
+ (\bar\psi_L\omega_R)(\bar\omega_R\psi_L) \right] + \cdots ,}
comes from the Fierz rearrangement of the operator corresponding to the
exchange of a massive coloron between the $\psi_L$ and the $\chi_R$ or
$\omega_R$ currents.  Other operators, such as those originating from $LL$
and $RR$ currents are suppressed in the $1/N_c\to 0$ limit.  This
interaction preserves a Peccei-Quinn symmetry.  In order to give the
``axion'' an acceptable mass, we add an explicit breaking term to the
Lagrangian:
\eqn\higgsaa{ \eqalign{ {\cal L}_{\rm int} 
&= {g_{\rm tc}^2\over M^2} \left[
(\bar\psi_L\chi_R)(\bar\chi_R\psi_L) 
+ (\bar\psi_L\omega_R)(\bar\omega_R\psi_L) \right] \cr
&\quad + \xi {g_{\rm tc}^2\over M^2} \left[
\epsilon_{ab} (\bar\psi^a_L\chi_R) (\bar\psi^b_L\omega_R)
+ \epsilon^{ab} (\bar\chi_R\psi_{aL}) (\bar\omega_R\psi_{bL}) \right]
+ \cdots.\cr} }
Here $\epsilon^{ab}$ is completely antisymmetric with $\epsilon^{12} = -
\epsilon^{21} = 1$.  We should point out that the Peccei-Quinn term we 
have added is only one of many possible terms that could arise from the
higher energy flavor physics.  We expect that this term should be small
compared to $\higgsa$, which translates into the requirement $\xi\ll 1$.
This follows from the fact that interactions originating from the flavor
physics are typically suppressed by the ratio of the topcolor scale ($\sim
10\, {\rm TeV}$) to the flavor scale ($\sim 100\, {\rm TeV}$).  We shall
show that both these conditions can be simultaneously met---that it is
possible to have $\xi$ small and the mass of the light pseudoscalar Higgs
well above current experimental bounds.

If we combine the $\chi_R$ and $\omega_R$ fields into a doublet,
\eqn\higgsb{ \lambda_R \equiv \pmatrix{\chi_R\cr \omega_R\cr} }
then the interaction Lagrangian becomes
\eqn\Lint{ {\cal L}_{\rm int} = {g_{\rm tc}^2\over M^2} 
(\bar\psi_{aL}\lambda^b_R)(\bar\lambda_{bR}\psi^a_L) 
+ {\xi\over 2} {g_{\rm tc}^2\over M^2} \left[
\epsilon^{ab} \epsilon_{cd}
(\bar\psi_{aL}\lambda^c_R) (\bar\psi_{bL}\lambda^d_R)
+ \epsilon^{ab} \epsilon_{cd}
(\bar\lambda_{aR}\psi^c_L) (\bar\lambda_{bR}\psi^d_L) \right].}

We next introduce a static, auxiliary field $\M^a_b$, which becomes a pair
of Higgs doublets in the low energy effective theory once we have
integrated out the heavy $\chi$ and $\omega$ fermions.  We introduce $\M$ 
through the effective Lagrangian 
\eqn\higgsc{ \eqalign{ {\cal L}_{\rm eff} 
&= g_{\rm tc} \left[ \bar\psi_{aL}\M^a_b\lambda^b_R
+ \bar\lambda_{aR}\M^{\dagger a}_b\psi^b_L
+ \xi \bar\lambda_{aR}\Mt^a_b\psi^b_L
+ \xi \bar\psi_{aL}\Mt^{\dagger a}_b\lambda^b_R \right] \cr
&\quad
- M^2 \left[ \Tr  (\M^\dagger\M) 
+ {\textstyle{1\over 2}} \xi\Tr  (\Mt\M)
+ {\textstyle{1\over 2}} \xi\Tr  (\Mt^\dagger\M^\dagger) \right].\cr} }
Note that the equations of motion for this static field,
\eqn\higgsd{ \eqalign{
\M^b_a &= {g_{\rm tc}\over M^2} \bar\psi_{aL}\lambda^b_R \cr
\Mt^b_a &= {g_{\rm tc}\over M^2} \epsilon^{bc}\bar\psi_{cL}\lambda^d_R 
  \epsilon_{da}
= \epsilon^{bc} \M^d_c \epsilon_{da}, \cr} }
when substituted back into the effective Lagrangian, reproduce
the original interaction Lagrangian, $\Lint$.  In the low energy theory,
we should integrate out the heavy degrees of freedom, which promotes the
auxiliary static field to a fully dynamic Higgs field.  To one-loop 
order, this procedure produces a propagating, self-interacting Higgs
field---working to first order in $\xi$, we find
$$\eqnn\oneloop
\eqalignno{ {\cal L}_{\rm 1-loop} 
&= Z_\phi\Tr(\partial_\mu\M^\dagger\partial^\mu\M)
+ \xi Z_\phi \left[ \Tr(\partial_\mu\Mt\partial^\mu\M)+\hbox{h.c.}\right]&\cr
&\quad
+ Z_m\Tr(\M^\dagger\M)
+ \xi \left[ Z_m\Tr(\Mt\M) + \hbox{h.c.} \right] &\cr
&\quad
- M^2 \left[ \Tr(\M^\dagger\M) 
+ {\textstyle{1\over 2}} \xi\Tr(\Mt\M)
+ {\textstyle{1\over 2}}\xi\Tr(\Mt^\dagger\M^\dagger) \right] &\oneloop \cr
&\quad - \lambda \biggl[ 
\Tr(\M^\dagger\M\M^\dagger\M)
+ \xi \left[ \Tr(\M^\dagger\M)\Tr(\Mt\M) + \hbox{h.c.} \right]
\biggr]. &\cr}$$
where 
\eqn\higgse{ \eqalign{
Z_\phi &\equiv {g_{\rm tc}^2N_c\over (4\pi)^2} \ln{M^2\over\mu^2} \cr
Z_m &\equiv {2g_{\rm tc}^2N_c\over (4\pi)^2}\left( M^2 - \mu^2 \right) \cr
\lambda &\equiv {g_{\rm tc}^4N_c\over (4\pi)^2} \ln{M^2\over\mu^2} . \cr} }
Observe that at energies $\mu<M$, below the scale at which the effective
field theory description breaks down, we have a fully dynamic Higgs field;
at the boundary, $\mu=M$, the one-loop effects that produce these dynamics
are small.  

We next evaluate the Higgs spectrum in the two-doublet model by expanding
the fields about the vacuum,\foot{This corresponds to the
case $\tan\beta=1$ in the notation of \ref\hhg{J.~F.~Gunion, H.~E.~Haber,
G.~Kane, and S.~Dawson, {\it The Higgs Hunter's Guide,\/} Addison-Wesley
Publishing Company, 1990.}.}
\eqn\vev{ \M = {v\over\sqrt{2}} \pmatrix{1 &0\cr  0 &1\cr}
\qquad\qquad \hbox{for } v=246\, {\rm GeV}/\sqrt{2} = 174\, {\rm GeV}. }
The fact that the two entries on the diagonal are equal is a consequence of
the $\M\leftrightarrow -\Mt$ symmetry of the Lagrangian, which we have
checked is left unbroken by the vacuum.  If we define the fields
representing fluctuations about this vacuum state by
\eqn\higgsm{ \eqalign{
\M^1_1 &= {\textstyle{1\over\sqrt{2}}}\left( v+H^0+h^0+iA^0+iG^0\right)\cr
\M^1_2 &= {\textstyle{1\over\sqrt{2}}} \left( H^+ - G^+ \right) \cr
\M^2_1 &= {\textstyle{1\over\sqrt{2}}} \left( H^- + G^- \right) \cr
\M^2_2 &= {\textstyle{1\over\sqrt{2}}}\left ( v+H^0-h^0+iA^0-iG^0\right),\cr} }
then we find that the fields $G^0$, $G^\pm$ are the Goldstone bosons
associated with the $SU(2)_W\times U(1)_Y\to U(1)_{\rm em}$ symmetry
breaking while the others represent a pair of neutral Higgs fields, $h^0$
and $H^0$, a charged Higgs field $H^\pm$, and a neutral pseudoscalar Higgs
field, $A^0$.  In the usual notation for two doublet models $\hhg$, these
linear combinations correspond to mixing angles $\alpha=-\pi/4$ and
$\beta=\pi/4$.  To leading order in the parameter $\xi$, inserting the
values from equation $\oneloop$ and canonically normalizing the fields, we
find that the masses of the $h^0$, $H^0$ and $H^\pm$ are degenerate,
\eqn\higgsq{ m^2_{h^0} = m^2_{H^0} = m^2_{H^\pm} \equiv m^2_H 
= {2v^2\lambda_1\over Z^2_\phi} 
= {32\pi^2 g_{\rm tc}^2 v^2\over N_c \ln(M^2/\mu^2)} . }
The pseudoscalar Higgs has a mass scaled down by a factor $\sqrt{2\xi}$,
\eqn\higgsr{ m^2_{A^0} = 2\xi m^2_H. }
The current lower bound on the mass of a pseudoscalar Higgs is about $62\,
{\rm GeV}$ \ref\pdg{Particle Data Group, ``Review of Particle Physics,''
Eur.~Phys.~J.~{\bf C3} (1998) 1.}, so that for $m_H\sim 1\, {\rm TeV}$,
even a small amount of Peccei-Quinn symmetry breaking, larger than $\xi
\sim 1/500$, is sufficient to be in accord with observations.  Such a value
of $\xi$ could easily be generated by physics at the flavor scale; if this
scale is an order of magnitude above the scale at which topcolor breaks, we
might expect $\xi\sim 10^{-2}$.

\newsec{Experimental Constraints on Topcolor Models.}

Most of the new physics in a topcolor model appears at scales of $1$--$10\,
{\rm TeV}$ or higher, so that the new fields, whether the heavy fermions or
the extra Higgs fields, can not be directly seen in current experiments.
However, their presence should appear in precise tests of the electroweak
theory, particularly in measurements of the $\rho$ parameter or $R_b$, the
ratio of decay width of $Z\to b\bar b$ to that of $Z\to\hbox{hadrons}$.  In
the one-doublet model, we use the effect of the new heavy fermion on Peskin
and Takeuchi's \peskin\  $T$ parameter combined with the Higgs's contribution
to $S$ and $T$ to map out the experimentally allowed region of the $m_{\rm
higgs}$--$m_\chi$ plane.  The allowed $\chi$ mass depends on the mass of
the Higgs, but for a $0.5$--$1\, {\rm TeV}$ Higgs, the 90\% confidence
level limits place $m_\chi$ between about $5$--$8\, {\rm TeV}$.  In the
two-doublet model, the limits on $R_b$ exclude an $\omega$ mass less than
than about $12\, {\rm TeV}$.  The mass of the $\chi$ is not so tightly
constrained.  Again, determining the contributions of the new physics, the
$\chi$, $\omega$ and new Higgs fields, to $S$ and $T$, we plot the allowed
regions in the $m_\chi$--$m_\omega$ planes for different choices of the
Higgs fields' masses.

To generate these plots for the allowed masses, we must first determine how
the new fermions contribute to $T$.  As mentioned in the introduction, one
of the advantages of choosing $\chi$ and $\omega$ to be weak singlets is
that they do not then contribute significantly to $S$.  We could analyze
$T= \alpha_{_{\rm QED}} \delta\rho\equiv \alpha_{_{\rm QED}}
(\rho-\rho_{\rm sm})$\foot{Here $\rho_{\rm sm}$ is the
standard model prediction to the $\rho$ parameter.} by summing the one-loop
vacuum polarization graphs for the $Z^0$ and $W^\pm$ propagators that
contain $\chi$ and $t$ fermion loops to find,
\eqn\drexact{ \eqalign{ \delta\rho
& = {N_c\over 16\pi^2 v^2} \left[ \sin^4\theta_L^\chi m^2_\chi +
2\sin^2\theta_L^\chi\cos^2\theta_L^\chi {m^2_\chi m^2_t\over m^2_\chi - m^2_t}
\ln {m^2_\chi\over m^2_t} 
- \sin^2\theta_L^\chi (2-\sin^2\theta_L^\chi) m_t^2 \right]
\cr} }
for the one-doublet theory and, including the $\omega$ and $b$ loops as
well,
\eqn\expa{ \eqalign{ \delta\rho
&= {N_c\over 16\pi^2 v^2} \Bigl[
\sin^4\theta_L^\chi f(m_\chi,m_t) + \sin^4\theta_L^\omega f(m_\omega,m_b) \cr
&\qquad
+ \sin^2\theta_L^\chi 
  \left( f(m_\chi,m_b) - f(m_t,m_b) - f(m_\chi,m_t) \right) \cr
&\qquad
+ \sin^2\theta_L^\omega 
  \left( f(m_\omega,m_t) - f(m_t,m_b) - f(m_\omega,m_b) \right) \cr
&\qquad
+ \sin^2\theta_L^\chi \sin^2\theta_L^\omega 
  \left( f(m_\chi,m_\omega) + f(m_t,m_b) 
       - f(m_\omega,m_t) - f(m_\chi,m_b) \right) \Bigr] \cr} }
for the two-doublet theory, where
\eqn\expb{ f(m_1,m_2) = m_1^2 + m_2^2 
             - {2m_1^2m_2^2\over m_1^2-m_2^2}\ln{m_1^2\over m_2^2}. }
Notice that $f(m_1,m_2)$ vanishes when $m_1=m_2$.  Also, $\theta^\chi_L$ is
the mixing angle between $\chi_L$ and $t_L$ that rotates these states into
the mass eigenstate basis, and similarly for
$\theta^\chi_R,~\theta^{\omega}_{L,R}$.

The calculations which led to these results are lengthy and provide little
insight into the physics so we shall extract the leading behavior via an
effective operator approach, described in 
\ref\grinwise{B.~Grinstein and M.~Wise, ``Operator analysis for precision 
electroweak physics,'' Phys.\ Lett.\ {\bf B265} (1991) 326.}.
Our calculations in this section do not include effects of the Peccei-Quinn
symmetry breaking term in $\higgsaa$.  We showed that these effects could
be small and still produce a mass for the pseudoscalar Higgs, and moreover
we expect them to be small when they originate from the flavor physics
since such effects are generically suppressed by powers of the ratio of the
topcolor scale to the flavor scale.  It is important, however, that
$\tan\beta\equiv v_2/v_1 \approx 1$, where $v_{1,2}$ are the vacuum
expectation values for the two Higgs doublets, since when the custodial
$SU(2)$ symmetry in the Higgs sector is broken, $\delta\rho$ receives
potentially large corrections that scale quadratically in the Higgs masses
\ref\toussaint{D.~Toussaint, ``Renormalization Effects from Superheavy
Higgs Particles,'' Phys.\ Rev.\ {\bf D18} (1978) 1626.}.  However, we have
explicitly checked that $\tan\beta =1$ is the minimum of the $\xi=0$ vacuum
so we set $\tan\beta=1$ in the following calculation for the two-doublet
model.

\subsec{The One-Doublet Model.}

Generically, the presence of new physics at some high energy scale $M\gg v$
appears at low energies in the form of non-renormalizable corrections to
the standard model.  Since these non-renormalizable operators arise when we
integrate out the heavy fields, they enter the low energy theory suppressed
by powers of $1/M$.  We can use the effective operator approach of 
\grinwise\  and \ref\mitch{M.~Golden and L.~Randall, ``Radiative
Corrections to Electroweak Parameters in Technicolor Theories,'' Nucl.\
Phys.\ {\bf B361} (1991) 3--23.}  to estimate the corrections to
$\delta\rho$ due to integrating out the $\chi$ and $\omega$ fields, so that
the relevant mass scale for $M$ is $m_\chi$ or $m_\omega$.  The operators
that produce the leading contribution to $\delta\rho$ involve four Higgs
fields, at most two derivatives and must break the custodial $SU(2)$
symmetry.  The only such operator is\foot{Here we have
labeled the operator to agree with the notation in \grinwise .}
\eqn\grinb{ {c_4\over m_\chi^2} {\cal O}_4 = {c_4\over m_\chi^2} 
(H^\dagger_a D_\mu H^a) (H^\dagger_b D^\mu H^b) + {\rm h.c.} }
More generally, a one-loop graph in the full theory will contain custodial
$SU(2)$-conserving parts as well; these can contribute to the effective
operators
\eqn\grinc{ {c_5\over m_\chi^2} {\cal O}_5 = {c_5\over m_\chi^2} 
(H^\dagger_a D^2 H^a) (H^\dagger_b H^b)  + {\rm h.c.} \qquad 
{c_6\over m_\chi^2} {\cal O}_6 = {c_6\over m_\chi^2} 
(D_\mu H^\dagger_a D^\mu H^a) (H^\dagger_b H^b). }

Let us first determine the matching contribution to ${\cal O}_4$ in the one
Higgs doublet model which comes from the one-loop diagram in figure 5.
Expanding this graph in powers of the external momenta and retaining the
quadratic terms, we find
\eqn\expd{ \cdots - {N_c\over 16\pi^2} \delta_d^a \delta_b^c
g_{\rm tc}^4\cos^4\theta^\chi_L \cos^4\theta^\chi_R 
\left[ {1\over 12} {1\over m_\chi^2} \left[ 3s + 9t +u \right] \right]
+ \delta_b^a \delta_d^c (s\leftrightarrow u) \cdots }
where $s$, $t$ and $u$ are the usual Mandelstam variables.  Matching this
to the effective theory, equations $\grinb$ and $\grinc$, 
\eqn\expe{ {i\over m_\chi^2} \delta_d^a \delta_b^c 
\left(  c_4\, {\textstyle{1\over 2}} (t-s-u)  + c_5\, (s+t+u) 
+ c_6\, {\textstyle{1\over 2}} (u-s-t) \right) 
+ {i\over m_\chi^2} \delta_b^a \delta_d^c (s\leftrightarrow u) }
we discover that the custodial $SU(2)$ violating piece is
\eqn\expf{ c_4 =  - {N_c\over 16\pi^2} {g_{\rm tc}^4\over 4}, }
upon taking $\theta^\chi_L \approx 0 \approx \theta^\chi_R$, which we
require for the top quark seesaw.  This leads to a shift in the $\rho$ 
parameter of \grinwise ,
\eqn\expg{ \delta\rho = - {c_4}{v^2\over m_\chi^2} = 
{N_c\over 16\pi^2} {g_{\rm tc}^4\over 4} {v^2\over m_\chi^2}, }
which agrees with the leading piece of the result of the exact, but much
lengthier, analysis of the one-loop vacuum polarization diagrams of
equation $\drexact$.

A second, logarithmic, contribution to $\delta\rho$ arises from running
from $\mu=m_\chi$ down to $m_t$.  In the theory below the scale of the
heavy fermions, integrating out the $\chi$ produces an effective operator
of the form,
\eqn\exph{ {\cal O} = {g^2_{\rm tc}\over m_\chi^2} 
(\bar\psi^3_{La} H^a) \gamma^\mu D_\mu (H^\dagger_b \psi^{3b}_L). }
When inserted into the diagrams shown in figure 6 (there represented by a
heavy dot), the piece of these diagrams which is quadratic in the momenta
($\lambda_t$ is the top quark Yukawa coupling),
\eqn\expj{ -{N_c\over 16\pi^2} {\lambda_t^2 g_{\rm tc}^2\over 2}
{\mu^\epsilon\over\epsilon} 
\delta_d^a \delta_b^c \left[ s + 3t - u \right] + \cdots , }
produces a contribution to $c_4$ given by,
\eqn\expk{ -{N_c\over 16\pi^2} {\lambda_t^2 g_{\rm tc}^2\over 2} 
\ln{\mu^2} + \cdots. }
Running between $m_\chi$ and $m_t$ gives a logarithmic correction to the
matching term found before, so that
\eqn\expl{ \delta\rho = 
{N_c\over 16\pi^2} {g_{\rm tc}^4\over 4} {v^2\over m_\chi^2} 
\left[ 1 + 2 {\lambda_t^2\over g_{\rm tc}^2} \ln{m_\chi^2\over m_t^2} 
\right], }
which agrees with the corresponding leading terms of the exact analysis to
leading order in $\theta^{\chi}$.  To see this, recall that
$\theta^\chi\simeq \lambda_t / g_{tc}$ for small $\lambda_t \ll g_{tc}$.

The allowed values for the $\chi$ mass for a range of Higgs masses are
summarized in figure 7.  In this figure we have shown the regions in the
$m_{\rm higgs}$--$m_{\chi}$ plane which agree with the latest set of
precision electroweak tests $\ewwg$\ to within a 68\% and a 90\% confidence
level.  The Higgs field makes its own contribution to the $S$ and $T$ (or
$\rho=\alpha_{_{\rm QED}} T$) parameters through $\peskin$
$$\eqalign{
S_{\rm higgs} &\approx {1\over 12\pi} 
              \ln{m^2_{\rm higgs}\over m^2_{\rm higgs,\ ref}}\cr
T_{\rm higgs} &\approx -{3\over 16\pi\cos^2\theta_W} 
              \ln{m^2_{\rm higgs}\over m^2_{\rm higgs,\ ref}}, \cr}
\neqno\sthiggs$$
where the reference Higgs mass was chosen to be $m_{\rm higgs,\ ref}=300\,
{\rm GeV}$.  This reference mass was also used in the ZFITTER 
\ref\zfitter{
D.~Bardin {\it et al.},  ``ZFITTER: An Analytical program for fermion 
pair production in $e^+ e^-$ annihilation,'' CERN-TH.\ 6443/92; 
{\tt hep-ph/9412201}.}
routine to obtain the Standard model estimates of the electroweak
parameters used to generate figure 7.  The plot essentially involves only
two parameters, $m_{\rm higgs}$ and $m_\chi$.  Therefore, we have used the
68\% and 90\% confidence levels for two degrees of freedom and added these
to the best fit value, $\chi^2=21$, which occurs for $m_{\rm
higgs}=159\, {\rm GeV}$ and $m_\chi\to\infty$.  This procedure actually
provides a conservative estimate for the allowed region.  If we use
instead the 68\% and 90\% confidence levels appropriate for the 19
parameters used to generate the standard model contributions to figure 7,
the allowed region expands slightly although the qualitative shape remains
unaltered.  In any event, for a relatively heavy Higgs field, $0.5$--$1.0\,
{\rm TeV}$, the mass of the $\chi$ fermion should lie between $5$ and $8\,
{\rm TeV}$.

Notice that as the $\chi$ mass grows large and its contribution to
$T=\alpha^{-1}_{_{\rm QED}}\delta\rho$ diminishes, the acceptable values
for $m_{\rm higgs}$ approach the usual range quoted in
\ref\pdghiggs{See for example the references cited in the table $H^0$ {\bf
Indirect Mass Limits from Electroweak Analysis} on page 247 of 
Particle Data Group, ``Review of Particle Physics,'' 
Eur.~Phys.~J.~{\bf C3} (1998) 1.};
for example, we find that to the 90\%  confidence level the mass range
which best fits the current electroweak data \ewwg\  for $m_\chi\to\infty$ is 
$$m_{\rm higgs} = 159^{+86}_{-56}\, {\rm GeV}. \neqno\nochi$$
The presence of a heavy fermion with a mass of $5$--$10\, {\rm TeV}$
completely alters these bounds, which do not include the effects of the
new physics, so that the Higgs can be as heavy as a ${\cal O}(1\, {\rm
TeV})$ while $S$ and $T$ still lie within the 90\% confidence level
region.  We should point out that this best fit value for the Higgs is
larger than the $76\, {\rm GeV}$ quoted in $\ewwg$.  This difference
can be explained by our choice of $\alpha_{_{\rm S}}(M_Z) = 0.118$ and 
$m_{\rm top}= 173.8\, {\rm GeV}$ as inputs for ZFITTER rather than the
values $\alpha_{_{\rm S}}(M_Z) = 0.119$ and $m_{\rm top}= 171.1\, {\rm
GeV}$ listed in table 32 of $\ewwg$, and by the fact that we are including only
the leading logarithmic dependence on the Higgs mass.  This is a rather
poor approxmation for small Higgs masses, but since our fits generally favor a
heavy Higgs, it is sufficient for our purposes.

\subsec{The Two-Doublet Model.}

Since the Higgs and its mass have yet to be observed, the precision
electroweak data do not sharply constrain the masses of the new fermions in
the two-doublet model.  The current measurements set lower bounds of about
$2.5\, {\rm TeV}$ for the $\chi$ fermion and about $12\, {\rm TeV}$ for the
$\omega$.  The reason for the higher bound on the $\omega$ mass is that
through mixing with the $b$ quark, it directly affects the prediction for
$R_b$ which has been precisely measured at LEP and SLD.  In contrast, the
$\chi$ mass affects the $T$ parameter along with the $\omega$ and Higgs
masses so that the experimental bounds on $m_\chi$ depend greatly upon the
particular values of the $\omega$ mass and the mass of the Higgs fields.
Paralleling our discussion for the one-doublet model, we first develop an
effective operator description for the contributions to $\delta\rho$.  In
the limit that the Peccei-Quinn breaking terms are small, the vacuum
expectation values of the two Higgs fields are equal, so we perform our
analysis with $\tan\beta=1$.  Operators that could arise from the
flavor-scale physics can generically perturb the vacuum away from
$\tan\beta=1$, but since such effects depend on the details of the flavor
physics, we only briefly consider $\delta\rho$ for $\tan\beta\not=1$
without attempting to estimate $\beta$.  After deriving the matching and
running contributions to $\delta\rho$, we derive a more stringent bound on
the $\omega$ mass by studying $Z\to b\bar b$.

As in the one-doublet model, we can write down the relevant dimension-six
custodial SU(2) violating operators,
$$\eqalign{
{\cal O}_4^{\chi} &= c_4^\chi (H_\chi^\dagger D_\mu H_\chi)
                              (H_\chi^\dagger D^\mu H_\chi) \cr
{\cal O}_4^{\omega} &= c_4^\omega (H_\omega^\dagger D_\mu H_\omega)
                                  (H_\omega^\dagger D^\mu H_\omega) \cr
{\cal O}_4^{\chi\omega} &= c_4^{\chi\omega}[(H_\omega^\dagger D_\mu H_\chi)
(H_\chi^\dagger D^\mu H_\omega) + \hbox{h.c.}] . \cr}
\neqno\grintdb$$
where $H_\chi^a={\cal M}^a_1$ and $H_\omega^a={\cal M}^a_2$, in terms of
our earlier matrix notation.\foot{We are neglecting the
Peccei-Quinn breaking terms which are small ($\xi=0$).}  The first two
operators are of the same form encountered in the one-doublet model, so we
can simply write down the one-loop matching contributions:
$$c_4^\chi = - {N_c\over 16\pi^2} {g_{\rm tc}^4\over 4}{1\over m_\chi^2} 
\qquad\hbox{and}\qquad
c_4^\omega = - {N_c\over 16\pi^2} {g_{\rm tc}^4\over 4}{1\over m_\omega^2} .
\neqno\expm$$
The third operator in equation $\grintdb$ partially undoes the effects of
the first two when $m_\chi\approx m_\omega$.  The leading contribution to
${\cal O}_4^{\chi\omega}$ from the full theory originates in the graph
shown in figure 8.  Just as in the one-doublet case, we must take care to
extract only the custodial-$SU(2)$ violating part of this graph.  In
addition to ${\cal O}_4^{\chi\omega}$, we can also write the following
dimension-six operators that contain two derivatives and two factors of
both $H_\chi$ and $H_\omega$:
$$\eqalign{ 
&c_5^{\chi\omega} [(H_\omega^\dagger D^2 H_\chi)(H_\chi^\dagger H_\omega) 
                 + \hbox{h.c.}] \cr
&\tilde c_5^{\chi\omega} 
   [(H_\omega^\dagger H_\chi)(H_\chi^\dagger D^2 H_\omega) + \hbox{h.c.}] \cr
&c_6^{\chi\omega}
   [(D_\mu H_\omega^\dagger D^\mu H_\chi)(H_\chi^\dagger H_\omega) 
   + \hbox{h.c.}], \cr}
\neqno\expo$$
which describe a complete set up to integrations by parts.  Retaining just
the $SU(2)$ violating piece, we find that
$$c_4^{\chi\omega} = - {N_c\over 16\pi^2} 
{g_{\rm tc}^4\over 4} {2\over m_\chi^2-m_\omega^2} 
\ln{m_\chi^2\over m_\omega^2}. \neqno\expp$$
The net matching contribution for the $\rho$-parameter due to the presence
of the heavy fermions is
$$\eqalign{\delta\rho 
&= -(c_4^\chi + c_4^\omega - c_4^{\chi\omega})v^2 \cr
&= {N_c\over 16\pi^2} {g_{\rm tc}^4 v^2\over 4}
\left[ {1\over m_\chi^2} + {1\over m_\omega^2} 
- {2\over m_\chi^2-m_\omega^2} \ln{m_\chi^2\over m_\omega^2} \right] .\cr}
\neqno\expq$$
Note that the operator ${\cal O}_4^{\chi\omega}$ contributes to
$\delta\rho$ with the opposite sign of ${\cal O}_4^\chi$ and ${\cal
O}_4^\omega$.  The origin of this sign can be seen when we write the vacuum
expectation values for the two Higgs fields as $\langle H_\chi\rangle =
\pmatrix{ v\cr 0\cr} $ and $\langle H_\omega\rangle = \pmatrix{0\cr
v\cr}$, where $v=246~{\rm GeV}/2=123$ GeV.  Then both ${\cal O}_4^\chi$ and
${\cal O}_4^\omega$ shift the $Z$ mass but leave the $W$ mass unaffected
while ${\cal O}_4^{\chi\omega}$ produces the opposite effect---it shifts
the $W$ mass while leaving the $Z$ mass unaltered.

After matching the full and effective theories at energies $\mu\approx
m_\chi, m_\omega$, $\delta\rho$ receives further logarithmic terms from
running down to energies, $\mu\approx m_t, m_b$.  The diagrams that produce
these running contributions involve insertions of the operators
$${g_{\rm tc}^2\over m_\chi^2} (\bar\psi_L H_\chi) D\!\!\!\!/ 
(H_\chi^\dagger\psi_L)
+ {g_{\rm tc}^2\over m_\omega^2} (\bar\psi_L H_\omega) D\!\!\!\!/ 
(H_\omega^\dagger\psi_L), \neqno\grintdrun$$
and resemble those encountered before for the one-doublet model, in figure
6.  Including these logarithmic corrections, we find the following form for
$\delta\rho$:
$$\eqalign{\delta\rho 
&= {N_c\over 16\pi^2} {g_{\rm tc}^4 v^2\over 4}
\Biggl[ {1\over m_\chi^2} + {1\over m_\omega^2} 
- {2\over m_\chi^2-m_\omega^2} \ln{m_\chi^2\over m_\omega^2}  \cr
&\quad\qquad\qquad
+{\lambda_t^2\over g_{\rm tc}^2} {2\over m_\chi^2}\ln{m_\chi^2\over m_t^2}
+{\lambda_b^2\over g_{\rm tc}^2} {2\over m_\omega^2}\ln{m_\omega^2\over m_b^2}
-{\lambda_b^2\over g_{\rm tc}^2} {2\over m_\chi^2}\ln{m_\chi^2\over m_b^2} 
-{\lambda_t^2\over g_{\rm tc}^2} {2\over m_\omega^2}\ln{m_\omega^2\over m_t^2} 
\Biggr] \cr} \neqno\expr$$
which reproduces the leading terms from the analysis $\expa$ of the
vacuum polarization graphs of the two Higgs doublet model.  Observe that in
the limit $m_\chi\to m_\omega$ that $\delta\rho\to 0$.

To complete our discussion of contributions to $\delta\rho$, we briefly
examine the case when $\tan\beta\not=1$.  Previously, we have assumed
that the vacuum expectation values for the two Higgs doublets were of equal
magnitude.  If effects arising from the flavor physics break this
equality, then we could have
$$\langle H_\chi\rangle = \pmatrix{v_1\cr 0\cr}\qquad\qquad
\langle H_\omega\rangle = \pmatrix{0\cr v_2\cr}, \neqno\higgsvevs$$
with 
$$v_1 \equiv v\cos\beta \qquad\qquad v_2 \equiv v\sin\beta,\neqno\polarvevs$$
where we have now reverted to the normalization where $v = 174$ GeV.
When $v_1\not= v_2$, then the estimate for $\delta\rho$ from the operators
that break the custodial $SU(2)$ becomes,
$$\delta\rho = -v^2 \left[ \cos^4\beta\, c^\chi_4 + \sin^4\beta\, c^\omega_4 
- \cos^2\beta\sin^2\beta\, c^{\chi\omega}_4 \right], \neqno\drhogen$$
or 
$$\delta\rho = {N_c\over 16\pi^2} {g_{\rm tc}^4v^2\over 4} 
\left[ {\cos^4\beta\over m_\chi^2} + {\sin^4\beta\over m_\omega^2}
- {2\cos^2\beta\sin^2\beta\over m_\chi^2 - m_\omega^2} \ln{m_\chi^2\over
m_\omega^2} \right] , \neqno\drhomatch$$
upon substituting in the matching contributions of equations
$\expm$ and $\expp$.  Even when the masses of the heavy fermions are equal,
$m_\chi=m_\omega$, the perturbation to the $\rho$-parameter need not
vanish:
$$\delta\rho = {N_c\over 16\pi^2} {g_{\rm tc}^4v^2\over 4} 
{\cos^2 2\beta\over m_\chi^2}. \neqno\drhomm$$

Returning to the $\tan\beta=1$ case, we can obtain a larger bound on the
mass of the $\omega$ by studying its effect on the ratio of decay widths
$R_b \equiv \Gamma[Z\to b\bar b]/\Gamma[Z\to\hbox{hadrons}]$.  The dominant
contribution to $R_b$ comes from the mixing of the $\omega$ and $b$ fields.
In passing from the weak eigenstates to the mass eigenstates, we perform a
rotation
\eqn\bmix{ \eqalign{
b_L &\to \cos\theta^\omega_L b_L - \sin\theta^\omega_L \omega_L \cr
b_R &\to \cos\theta^\omega_R b_R - \sin\theta^\omega_R \omega_R, \cr} }
which shifts the $Zb\bar b$ couplings slightly.
The standard model coupling of the $Z$ to the left-handed $b$ quark,
\eqn\exps{ g_L^{b\, \rm sm} = \left[ {\textstyle{1\over 3}}\sin^2\theta_W 
                   - {\textstyle{1\over 2}} \right], }
becomes after rotating the fields,
\eqn\expt{ g^b_L = g_L^{b\, \rm sm}\, \cos^2\theta^\omega_L + g_L^\omega\,
\sin^2\theta^\omega_L \approx g_L^{b\, \rm sm} 
+ {1\over 2} {m^2_{b\omega}\over m^2_\omega } , }
to leading order.  Here $m_{b\omega}$ is the dynamically generated mass
that results from $b_L \omega_R$ condensation.  Since $b_R$ and $\omega_R$
have the same couping to the $Z$, there is no shift in $g_R$.  In the NJL
approximation for a two doublet model, $m_{b\omega} = 400$ GeV.  The mixing
leads to a slight shift in the standard model prediction:
\eqn\expu{ R_b 
= R_b^{\rm sm} - 0.39 {m^2_{b\omega}\over m^2_\omega }. }
The standard model prediction, $R_b^{\rm sm}$, is $0.2157\pm 0.0004$
\ref\hill{C.~T.~Hill and X.~Zhang, ``$Z\to b\bar b$ versus Dynamical
Electroweak Symmetry Breaking Involving the Top Quark,'' Phys.\ Rev.\ {\bf
D51} (1995) 3563.} while the most recent measurements yield $R_b =
0.21656\pm 0.00074$ so that to agree to within $2\sigma$ requires 
\eqn\expv{
m_\omega  > 25 m_{b\omega} \sim 10\, {\rm TeV}.}

Apart from this tree-level mixing, the strongly coupled Higgs fields can
also modify the width of $Z\rightarrow b\bar b$ through one-loop effects.
We can estimate these effects using the effective theory given in
equation $\higgsc$,\foot{Here we have assumed that the
Higgs field has been renormalized so that we use the renormalized coupling
constant $g_{\rm t} = g_{\rm tc}/\sqrt{Z_\phi}$, where $Z_\phi$ is given in
equation $\higgse$.}
$${\cal L}_{\rm eff} = g_{\rm t} \bar\psi^3_L {\cal M} \lambda_R + {\rm h.c.}
\neqno\loopsa$$
We shall show that most of the one-loop effects are generally small so we 
have neglected the Peccei-Quinn terms which are additionally suppressed by
$\xi$.  If we rotate these interactions into a mass eigenstate basis, as in
equation $\bmix$ with an analogous pair of $t_{L,R}$--$\chi_{L,R}$
rotations, we find the following couplings:
$$\eqalign{
\bar b_L {\cal M}^2_1 \chi_R
   &: g_{\rm t} \cos\theta_L^\omega\cos\theta_R^\chi 
      \simeq {m_{t\chi}\over v} \cr
\bar b_L {\cal M}^2_1 t_R
   &: g_{\rm t} \cos\theta_L^\omega \sin\theta_R^\chi \simeq {m_t\over v} \cr
\bar b_L {\cal M}^2_2 \omega_R
   &: g_{\rm t} \cos\theta_L^\omega \cos\theta_R^\omega 
      \simeq {m_{b\chi}\over v} \cr
\bar b_L {\cal M}^2_2 b_R
   &: g_{\rm t} \cos\theta_L^\omega \sin\theta_R^\omega \simeq {m_b\over
      v}, \cr} \neqno\loopsb
$$
where the components of ${\cal M}$ can be expanded about the $\tan\beta=1$
vacuum according to equation $\higgsm$.  For $\tan\beta=1$, using the
Pagels-Stokar relation, we find $m_{b\omega}= m_{t\chi}\simeq 400$ GeV.
Yukawa couplings involving $\bar b_R \chi_L$, $\bar b_R t_L$, and $\bar
b_R\omega_L$ are all proportional to $\sin\theta_R^\omega\simeq m_b/v$ and
are therefore negligible.  From equation $\loopsb$, we conclude that the
dominant loop corrections are those shown in figures 9--11.

The diagrams of figure 9 have been studied previously 
\ref\dghk{A.~Denner, R.~J.~Guth, W.~Hollik, and J.~H.~Kuhn, ``The $Z$ width 
in the two Higgs doublet model,'' Z.\ Phys.\ {\bf C51} (1991) 695.  
A.~K.~Grant, ``Implications of a heavy top quark for the two Higgs 
doublet model,'' Phys.\  Rev.\  {\bf D51} (1995) 207; 
{\tt hep-ph/9410267}.}  
in the context of generic two-Higgs doublet models.  The correction is
negative and falls off as $1/m_{H^\pm}^2$ with increasing Higgs mass.  A
$300\, {\rm GeV}$ charged Higgs mass decreases $R_b$ by about one
$\sigma$.  In any case for a lighter $\chi$ mass, a heavy charged Higgs is
preferred and its effects will strengthen the bound $\expv$ on the $\omega$ 
mass.

The diagrams of figures 10--11 involve the heavy isosinglet fermions $\chi$
and $\omega$, both of which are strongly coupled to the composite Higgs
scalars.  We might therefore expect these corrections to be large.
However, since the $\chi$ and $\omega$ are vectorlike, they can be given
large $SU(2)_L$-invariant masses and must decouple in the large mass limit.
For this reason, the diagrams of figure 10 turn out to be quite small.  For
example, even with $m_{H^\pm} = 200\, {\rm GeV}$ and $m_\chi = 1\, {\rm
TeV}$, we find that these diagrams give
$$\delta R_b \simeq 2\times 10^{-5},$$
which is indeed negligible.  The neutral Higgs diagrams of figure 11 are
negative and typically quite small.  These corrections can be appreciable
if there is a large mass splitting between the $H^0$ and the $A^0$, in
which case the diagrams grow as $\ln(m_{H^0}^2/m_{A^0}^2)$ For
example, with $m_{A^0} = 100\, {\rm GeV}$, $m_{H^0} = 1 {\rm TeV}$, 
$m_{h^0} = 200\, {\rm GeV}$, and $m_\omega = 2\, {\rm TeV}$, we find
$$\delta R_b \simeq -0.001.$$
The neutral Higgs effects can shift $R_b$ by as much as one standard
deviation in extreme cases.  Finally, we note that the loop corrections
scale as $g_{\rm t}^2$, which is inversely proportional to
$\ln(M^2/m_\chi^2)$.  We have taken this logarithm to be 5,
corresponding to an order-of-magnitude difference between the topcolor
scale and the masses of the $\chi$ and $\omega$.  The results should be
scaled appropriately for models with different scales.

The corrections involving the Peccei-Quinn violating terms in the effective
Lagrangian are expected to be smaller simply because the coupling in this
case is weaker.  For present purposes it is sufficient to retain only the
Peccei-Quinn preserving part, with the caveat that the above results will
likely be modified at the few percent level by the Peccei-Quinn breaking
term.

In summary, despite the strong coupling between the weak-isosinglet
fermions and the Higgs fields, we find that loop corrections involving
these particles are typically small.  The loop effects from the graphs in
figure 9, however, do give a suppression of $R_b$ when the Higgs mass is
less than $500\, {\rm GeV}$, which will strengthen the lower bound $\expv$
on the mass of the $\omega$.

The Higgs fields themselves contribute to $S$
through\foot{The full contribution to $S$ from the Higgs
sector, with the standard model contributions subtracted out, is
$$\eqalign{
\Delta S &= {1\over 12\pi} \biggl[ 
\cos^2(\beta-\alpha) \ln{m_{H^0}^2\over m_{h^0}^2} - {11\over 6} \cr
&\quad\qquad
+ \sin^2(\beta-\alpha)\biggl(
{m_{H^0}^4+m_{A^0}^4\over (m_{H^0}^2-m_{A^0}^2)^2}
+ {(m_{H^0}^2-3m_{A^0}^2)m_{H^0}^4 \ln{m_{H^0}^2\over m_{H^\pm}^2}
  -(m_{A^0}^2-3m_{H^0}^2)m_{A^0}^4 \ln{m_{A^0}^2\over m_{H^\pm}^2}
     \over (m_{H^0}^2-m_{A^0}^2)^3}
\biggr) \cr
&\quad\qquad
+ \cos^2(\beta-\alpha)\biggl(
{m_{h^0}^4+m_{A^0}^4\over (m_{h^0}^2-m_{A^0}^2)^2}
+ {(m_{h^0}^2-3m_{A^0}^2)m_{h^0}^4 \ln{m_{h^0}^2\over m_{H^\pm}^2}
  -(m_{A^0}^2-3m_{h^0}^2)m_{A^0}^4 \ln{m_{A^0}^2\over m_{H^\pm}^2}
     \over (m_{h^0}^2-m_{A^0}^2)^3}
\biggr) \biggr] \cr}$$
where $\alpha$ and $\beta$ are defined as in $\hhg$.  The case we are
studying has $\beta= -\alpha = \pi/4$.}

$$\eqalign{
S_{\rm higgs} &= {1\over 12\pi} \biggl[ 
\ln{m_{h^0}^2\over m_{h^0, {\rm ref}}^2} 
+ {m_{H^0}^4+m_{A^0}^4\over (m_{H^0}^2-m_{A^0}^2)^2}
+ {(m_{H^0}^2-3m_{A^0}^2)m_{H^0}^4\over (m_{H^0}^2-m_{A^0}^2)^3}
  \ln{m_{H^0}^2\over m_{H^\pm}^2} \cr
&\quad\qquad
- {(m_{A^0}^2-3m_{H^0}^2)m_{A^0}^4\over (m_{H^0}^2-m_{A^0}^2)^3}
  \ln{m_{A^0}^2\over m_{H^\pm}^2} 
- {11\over 6} \biggr] \cr} \neqno\stwohiggs$$
and to $T$ through
\ref\dgk{A.~Denner, R.~J.~Guth and J.~H.~Kuhn,
``Relaxation Of Top Mass Limits In The Two Higgs Doublet Model,''
Phys.\ Lett.\ {\bf B240} (1990) 438.}
$$\eqalign{
T_{\rm higgs} &= {1\over 32\pi \sin^2\theta_W\cos^2\theta_W M_Z^2} \left[
f(m_{H^\pm},m_{A^0}) + f(m_{H^\pm},m_{H^0}) 
- f(m_{A^0},m_{H^0}) \right] \cr
&\quad - {3\over 16\pi \cos^2\theta_W} 
\ln{m_{h^0}^2\over m_{h^0, {\rm ref}}^2} \cr}
\neqno\ttwohiggs$$
where the function $f(m_1,m_2)$ was defined in equation $\expb$.  We have
used a reference value of $m_{h^0,\ {\rm ref}}=300\, {\rm GeV}$ in our
fits.  Combining the Higgs and $\chi$--$\omega$ contributions to $S$ and
$T$ and comparing to the current experimental constraints on these
parameters, we find the allowed values for $m_\chi$ and $m_\omega$, for
three illustrative Higgs masses, $m_{h^0}=400$, $800$ and $1200\, {\rm
GeV}$, shown in figure 12.  When the Peccei-Quinn breaking terms are
neglected, the masses of the heavy Higgs fields are degenerate, as seen in
equation $\higgsq$, so we have set $m_{h^0}=m_{H^0}=m_{H^\pm}$ in making
these plots.  We have also set $m_{A^0}=100\, {\rm GeV}$ to be safely above
experimental limits.

As we mentioned above, for $m_{h^0}\approx 300\, {\rm GeV}$, the
corrections from the loops shown in figure 9 decrease $R_b$ by one
$\sigma$.  In order to have $R_b$ lie within the experimentally acceptable
range then requires that the tree level corrections $\expu$ be small which
occurs when $m_\omega > 15\, {\rm TeV}$.  As we increase the common heavy
Higgs mass, the loop corrections become less important and we can permit a
larger tree level correction, which means that the bound on $m_\omega$ is
relaxed to about $12\, {\rm TeV}$ for $m_{h^0}\approx 800\, {\rm GeV}$.
When the Higgs masses are of the order of $1\, {\rm TeV}$ or heavier, then
the negative contribution to $T$ from the Higgs fields must be compensated
by a positive contribution from the heavy fermions, $\expr$.  This can only
occur if the $m_\chi=m_\omega$ symmetry is badly broken.  We see this
effect appearing in figure 12 where the allowed values for $m_\chi$ are in
the $3$--$5\, {\rm TeV}$ range while $m_\omega > 15\, {\rm TeV}$.

\newsec{Conclusions.}

We have presented a class of models that implement the top-condensate
see-saw mechanism of electroweak symmetry breaking.  The models accomplish
all gauge symmetry breaking dynamically, without recourse to
phenomenological Higgs scalars.  The gauge structure of the models is
complex.  This complexity results mostly from the requirement that the
models should admit higher dimensional gauge invariant operators that
generate fermion masses and Yukawa couplings in the low energy theory.
It may be possible to construct simpler models, once the flavor sector of
the theory is better understood.

The models we have considered involve new isosinglet quarks.  Models with
only one such quark yield a low energy theory with one Higgs doublet,
models with two yield two Higgs doublets.  As with most models of dynamical
symmetry breaking, the Higgs bosons are expected to be heavy.  In the two
doublet models, we expect a set of heavy ($\sim$ TeV) charged and neutral
scalars, together with a pseudoscalar which may be light ($\sim 100$ GeV).

The new isosinglet quarks have measurable effects on low-energy physics.
In one-doublet models, the heavy singlet $\chi$ can give a sizable
contribution to the $\rho$ parameter.  This contribution constrains the
mass of the $\chi$ to be between 5 and 8 TeV when the Higgs mass is above
500 GeV.  In two doublet models, the contribution to $\rho$ is small when
the $\chi$ and $\omega$ are degenerate, since in this case the model has an
$SU(2)_R$ custodial symmetry.  The most stringent constraint on two-doublet
models comes from $Z\rightarrow b\bar{b}$, which receives sizable
corrections from $b$-$\omega$ mixing.  We have shown that the $\omega$ must
at least be heavier than about 10 TeV.  Loop corrections, while small for
large Higgs masses and model dependent, make this bound much more stringent
when the Higgs masses are less than about $500\, {\rm GeV}$.  The loop and
tree level effects combine to impose a bound of $m_\omega > 12\, {\rm
TeV}$.

A number of interesting issues have yet to be resolved.  The models we have
presented serve to illustrate the main issues involved in topcolor
model-building.  The gauge symmetries of these models are typically quite
complex, and we might hope that a better understanding of the dynamics of
these models, together with a better understanding of the flavor structure,
will lead to simpler scenarios.  We hope to return to these questions in
later work.

\listrefs

\vfill\eject


$$\beginpicture
\setcoordinatesystem units <1.00truein,1.00truein>
\setplotarea x from -1.0 to 1.0, y from -0.70 to 0.70
\circulararc 360 degrees from 0.25 0 center at 0 0
\circulararc 360 degrees from 1.0 0 center at 1.25 0
\circulararc 360 degrees from -1.0 0 center at -1.25 0
\putrule from 0.25 0 to 1.0 0
\putrule from -1.0 0 to -0.25 0
\arrow <7pt> [0.2,0.67] from -0.545 0 to -0.705 0
\arrow <7pt> [0.2,0.67] from 0.705 0 to 0.545 0
\put {$SU(3)_1$} [c] at -1.25 0
\put {$SU(4)$} [c] at 0 0
\put {$SU(3)_2$} [c] at 1.25 0
\put {$\xi_R$} [c] at -0.625 0.15
\put {$\xi_L$} [c] at 0.625 0.15
\plot 1.491 0.065   2.216 0.259 /
\plot 1.491 -0.065  2.216 -0.259 /
\plot 1.427 0.177  1.957 0.707 /
\plot 1.427 -0.177  1.957 -0.707 /
\arrow <7pt> [0.2,0.67] from 1.491 0.065 to 1.931 0.182
\arrow <7pt> [0.2,0.67] from 1.427 0.177 to 1.749 0.499
\arrow <7pt> [0.2,0.67] from 2.216 -0.259 to 1.776 -0.141
\arrow <7pt> [0.2,0.67] from 1.957 -0.707 to 1.635 -0.385
\put {$\chi_L$} [l] at 2.00 0.70
\put {$\omega_L$} [l] at 2.25 0.25
\put {$U^{1,2,3}_R$} [l] at 2.25 -0.25
\put {$D^{1,2,3}_R$} [l] at 2.00 -0.70
\putrule from -1.5 0 to -2.25 0
\plot -1.467 0.125  -2.116 0.500 /
\plot -1.467 -0.125  -2.116 -0.500 /
\arrow <7pt> [0.2,0.67] from -2.116 0.500 to -1.722 0.273
\arrow <7pt> [0.2,0.67] from -2.25 0 to -1.795 0
\arrow <7pt> [0.2,0.67] from -1.467 -0.125 to -1.861 -0.353
\put {$\chi_R$} [r] at -2.15 0.50
\put {$\omega_R$} [r] at -2.30 0
\put {$\psi^{1,2,3}_L$} [r] at -2.15 -0.50
\endpicture$$
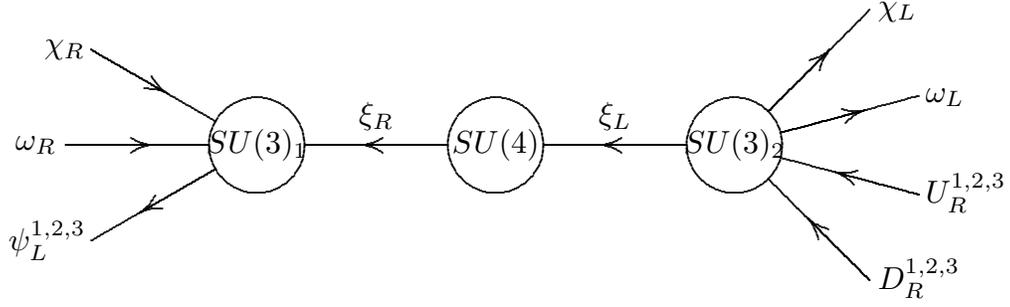
{\bf Figure 1.}  An example of a simple model that can realize a topcolor
seesaw.  The central gauge group is chosen so that the theory is free of
anomalies.
\vskip0.125truein

$$\beginpicture
\setcoordinatesystem units <1.00truein,1.00truein>
\setplotarea x from -2.0 to 2.0, y from -0.50 to 0.50
\circulararc 360 degrees from -1.75 0 center at -2.0 0
\circulararc 360 degrees from -0.75 0 center at -1.0 0
\circulararc 360 degrees from 0.25 0 center at 0 0
\circulararc 360 degrees from 0.75 0 center at 1.0 0
\circulararc 360 degrees from 1.75 0 center at 2.0 0
\putrule from -1.25 0 to -1.75 0
\putrule from -0.25 0 to -0.75 0
\putrule from 0.25 0 to 0.75 0
\putrule from 1.25 0 to 1.75 0
\arrow <7pt> [0.2,0.67] from -1.42 0 to -1.58 0
\arrow <7pt> [0.2,0.67] from -0.42 0 to -0.58 0
\arrow <7pt> [0.2,0.67] from 0.58 0 to 0.42 0
\arrow <7pt> [0.2,0.67] from 1.58 0 to 1.42 0
\put {$SU(3)$} [c] at -2 0
\put {$SU(5)$} [c] at -1 0
\put {$SU(3)$} [c] at 0 0
\put {$SU(6)$} [c] at 1 0
\put {$SU(3)$} [c] at 2 0
\put {$\xi_R$} [c] at -1.5 0.15
\put {$\xi_L$} [c] at -0.5 0.15
\put {$\zeta_R$} [c] at 0.5 0.15
\put {$\zeta_L$} [c] at 1.5 0.15
\putrule from 0 0.25 to 0 0.75
\putrule from 0 -0.25 to 0 -0.75
\arrow <7pt> [0.2,0.67] from 0 0.25 to 0 0.58
\arrow <7pt> [0.2,0.67] from 0 -0.75 to 0 -0.42
\put {$\psi^3_R$} [l] at 0.15 0.50
\put {$\chi_R$} [l] at 0.15 -0.50
\putrule from 2 0.25 to 2 0.75
\putrule from 2 -0.25 to 2 -0.75
\arrow <7pt> [0.2,0.67] from 2 0.75 to 2 0.42
\arrow <7pt> [0.2,0.67] from 2 -0.75 to 2 -0.42
\put {$U^{1,2,3}_R$} [l] at 2.15 0.50
\put {$D^{1,2,3}_R$} [l] at 2.15 -0.50
\putrule from -2 0.25 to -2 0.75
\putrule from -2 -0.25 to -2 -0.75
\arrow <7pt> [0.2,0.67] from -2 0.25 to -2 0.58
\arrow <7pt> [0.2,0.67] from -2 -0.25 to -2 -0.58
\put {$\chi_L$} [l] at -1.85 0.50
\put {$\psi^{1,2}_L$} [l] at -1.85 -0.50
\endpicture$$
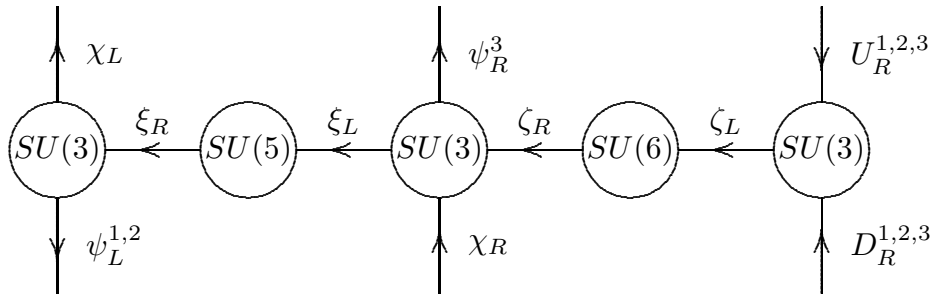
{\bf Figure 2.}  Another topcolor theory constructed so that $\bar\chi_L
t_R$ and $\bar\chi_R\chi_L$ terms are forbidden at tree-level.  Instead,
these terms arise dynamically in the low energy theory.
\vskip0.125truein

$$\beginpicture
\setcoordinatesystem units <1.00truein,1.00truein>
\setplotarea x from -1.5 to 1.5, y from -1.00 to 3.20
\circulararc 360 degrees from 0.25 0 center at 0 0
\circulararc 360 degrees from 1.0 0 center at 1.25 0
\circulararc 360 degrees from -1.0 0 center at -1.25 0
\putrule from 0.25 0 to 1.0 0
\putrule from -1.0 0 to -0.25 0
\arrow <7pt> [0.2,0.67] from -0.545 0 to -0.705 0
\arrow <7pt> [0.2,0.67] from 0.705 0 to 0.545 0
\put {$SU(3)_2$} [c] at -1.25 0
\put {$SU(m)$} [c] at 0 0
\put {$SU(3)_1$} [c] at 1.25 0
\put {$\xi_R^1$} [c] at -0.625 0.15
\put {$\xi_L^1$} [c] at 0.625 0.15
\putrule from 1.50 0 to 2.25 0
\plot 1.467 -0.125  2.116 -0.500 /
\plot 1.375 -0.217  1.750 -0.866 /
\arrow <7pt> [0.2,0.67] from 2.25 0 to 1.795 0
\arrow <7pt> [0.2,0.67] from 2.116 -0.500 to 1.722 -0.273
\arrow <7pt> [0.2,0.67] from 1.375 -0.217 to 1.603 -0.611
\put {$\chi_R$} [l] at 2.30 0
\put {$b_R$} [l] at 2.15 -0.50
\put {$\psi^3_L$} [l] at 1.80 -0.90
\putrule from -1.50 0 to -2.25 0
\plot -1.467 -0.125  -2.116 -0.500 /
\plot -1.375 -0.217  -1.750 -0.866 /
\arrow <7pt> [0.2,0.67] from -1.50 0 to -1.955 0
\arrow <7pt> [0.2,0.67] from -2.116 -0.500 to -1.722 -0.273
\arrow <7pt> [0.2,0.67] from -1.750 -0.866 to -1.523 -0.472
\put {$\psi^{1,2}_L$} [r] at -2.30 0
\put {$D^{1,2}_R$} [r] at -2.15 -0.50
\put {$U^{1,2,3}_R$} [r] at -1.80 -0.90
\circulararc 360 degrees from 0.875 1.083 center at 0.625 1.083
\circulararc 360 degrees from -0.875 1.083 center at -0.625 1.083
\circulararc 360 degrees from 0.25 2.165 center at 0 2.165
\put {$SU(3)'_1$} [c] at 0 2.165
\put {$SU(m)$} [c] at 0.625 1.083
\put {${\scriptstyle SU(m+1)}$} [c] at -0.625 1.083
\putrule from 0 2.415 to 0 3.165
\plot 1.125 0.217  0.750 0.866 /
\plot -1.125 0.217  -0.750 0.866 /
\arrow <7pt> [0.2,0.67] from -1.125 0.217 to -0.898 0.611
\arrow <7pt> [0.2,0.67] from 0.750 0.866 to 0.978 0.472
\put {$\xi_L^2$} [r] at -1.05 0.55
\put {$\xi_R^3$} [l] at  1.05 0.55
\putrule from 0 2.415 to 0 3.165
\arrow <7pt> [0.2,0.67] from 0 2.415 to 0 2.870
\put {$\chi_L$} [l] at 0.10 2.80
\plot 0.50 1.299  0.125 1.949 /
\plot -0.50 1.299  -0.125 1.949 /
\arrow <7pt> [0.2,0.67] from 0.125 1.949 to 0.353 1.555
\arrow <7pt> [0.2,0.67] from -0.50 1.299 to -0.273 1.693
\put {$\xi_L^3$} [l] at  0.45 1.60
\put {$\xi_R^2$} [r] at -0.45 1.60
\endpicture$$
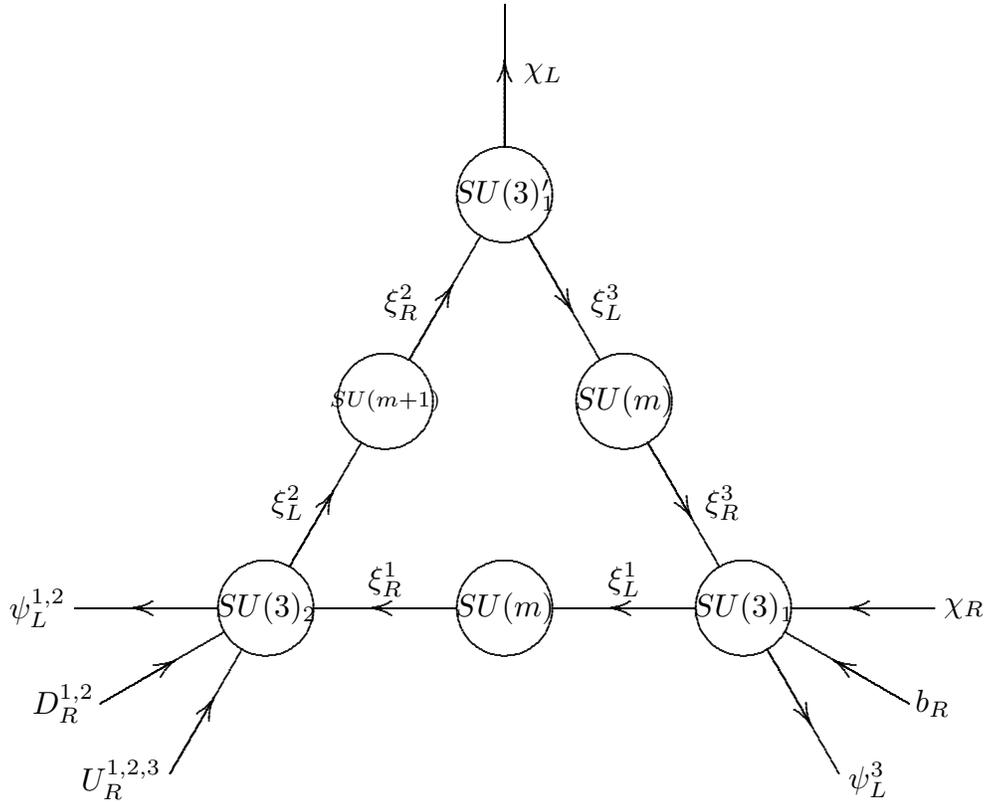
{\bf Figure 3.}  Chivukula, Dobrescu, Georgi, and Hill's topcolor model
$\seesaw$.  The low energy effective theory for this model contains a single
composite Higgs double.
\vskip0.125truein

$$\beginpicture
\setcoordinatesystem units <1.00truein,1.00truein>
\setplotarea x from -1.5 to 1.5, y from -1.00 to 3.20
\circulararc 360 degrees from 0.25 0 center at 0 0
\circulararc 360 degrees from 1.0 0 center at 1.25 0
\circulararc 360 degrees from -1.0 0 center at -1.25 0
\putrule from 0.25 0 to 1.0 0
\putrule from -1.0 0 to -0.25 0
\arrow <7pt> [0.2,0.67] from -0.545 0 to -0.705 0
\arrow <7pt> [0.2,0.67] from 0.705 0 to 0.545 0
\put {$SU(3)_2$} [c] at -1.25 0
\put {$SU(m)$} [c] at 0 0
\put {$SU(3)_1$} [c] at 1.25 0
\put {$\xi_R^1$} [c] at -0.625 0.15
\put {$\xi_L^1$} [c] at 0.625 0.15
\putrule from 1.50 0 to 2.25 0
\plot 1.467 -0.125  2.116 -0.500 /
\plot 1.375 -0.217  1.750 -0.866 /
\arrow <7pt> [0.2,0.67] from 2.25 0 to 1.795 0
\arrow <7pt> [0.2,0.67] from 2.116 -0.500 to 1.722 -0.273
\arrow <7pt> [0.2,0.67] from 1.375 -0.217 to 1.603 -0.611
\put {$\chi_R$} [l] at 2.30 0
\put {$\omega_R$} [l] at 2.15 -0.50
\put {$\psi^3_L$} [l] at 1.80 -0.90
\putrule from -1.50 0 to -2.25 0
\plot -1.467 -0.125  -2.116 -0.500 /
\plot -1.375 -0.217  -1.750 -0.866 /
\arrow <7pt> [0.2,0.67] from -1.50 0 to -1.955 0
\arrow <7pt> [0.2,0.67] from -2.116 -0.500 to -1.722 -0.273
\arrow <7pt> [0.2,0.67] from -1.750 -0.866 to -1.523 -0.472
\put {$\psi^{1,2}_L$} [r] at -2.30 0
\put {$D^{1,2,3}_R$} [r] at -2.15 -0.50
\put {$U^{1,2,3}_R$} [r] at -1.80 -0.90
\circulararc 360 degrees from 0.875 1.083 center at 0.625 1.083
\circulararc 360 degrees from -0.875 1.083 center at -0.625 1.083
\circulararc 360 degrees from 0.25 2.165 center at 0 2.165
\put {$SU(3)'_1$} [c] at 0 2.165
\put {$SU(m)$} [c] at 0.625 1.083
\put {${\scriptstyle SU(m+2)}$} [c] at -0.625 1.083
\plot 1.125 0.217  0.750 0.866 /
\plot -1.125 0.217  -0.750 0.866 /
\arrow <7pt> [0.2,0.67] from -1.125 0.217 to -0.898 0.611
\arrow <7pt> [0.2,0.67] from 0.750 0.866 to 0.978 0.472
\put {$\xi_L^2$} [r] at -1.05 0.55
\put {$\xi_R^3$} [l] at  1.05 0.55
\plot 0.125 2.382  0.500 3.031 /
\plot -0.125 2.382  -0.500 3.031 /
\arrow <7pt> [0.2,0.67] from 0.125 2.382 to 0.353 2.776
\arrow <7pt> [0.2,0.67] from -0.125 2.382 to -0.353 2.776
\put {$\chi_L$} [l] at 0.40 2.70
\put {$\omega_L$} [r] at -0.40 2.70
\plot 0.50 1.299  0.125 1.949 /
\plot -0.50 1.299  -0.125 1.949 /
\arrow <7pt> [0.2,0.67] from 0.125 1.949 to 0.353 1.555
\arrow <7pt> [0.2,0.67] from -0.50 1.299 to -0.273 1.693
\put {$\xi_L^3$} [l] at  0.45 1.60
\put {$\xi_R^2$} [r] at -0.45 1.60
\endpicture$$
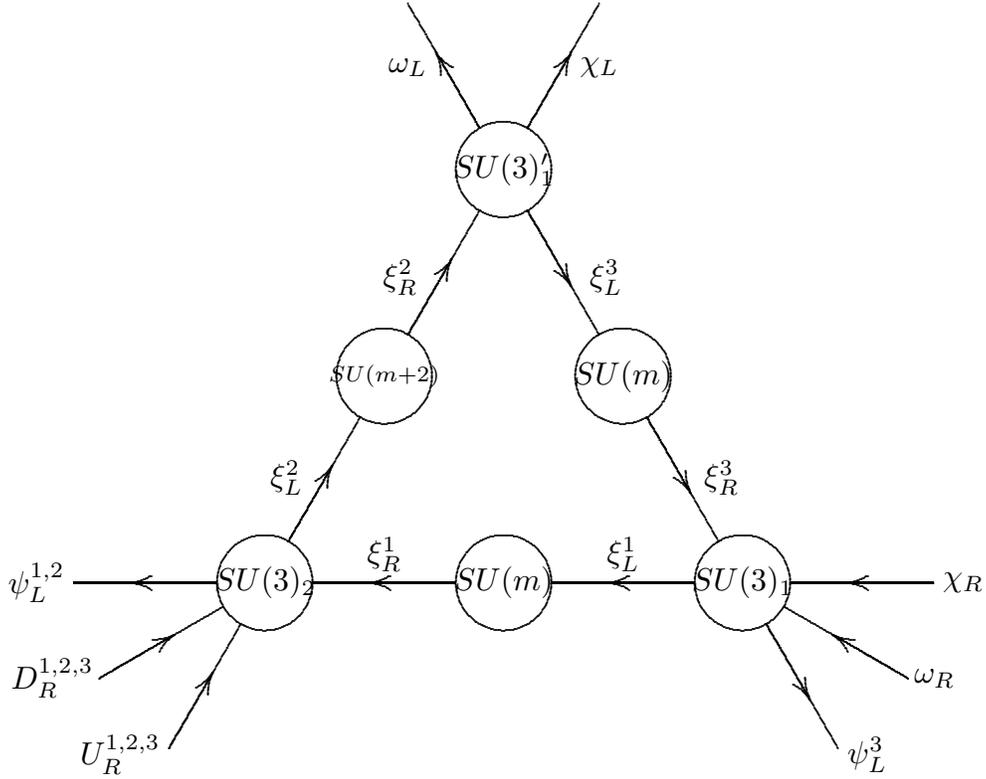
{\bf Figure 4.}  Another topcolor model which contains two composite Higgs
doublets in the low energy effective theory.
\vskip0.125truein

$$\beginpicture
\setcoordinatesystem units <1.00truein,1.00truein>
\setplotarea x from -1.5 to 1.5, y from 0.0 to 0.5
\putrule from 0.25 0.25 to -0.25 0.25 
\putrule from -0.25 0.25 to -0.25 -0.25 
\putrule from -0.25 -0.25 to 0.25 -0.25 
\putrule from 0.25 -0.25 to 0.25 0.25 
\arrow <7pt> [0.2,0.67] from 0.08 0.25 to -0.08 0.25
\arrow <7pt> [0.2,0.67] from -0.08 -0.25 to 0.08 -0.25
\arrow <7pt> [0.2,0.67] from 0.25 -0.08 to 0.25 0.08
\arrow <7pt> [0.2,0.67] from -0.25 0.08 to -0.25 -0.08
\plot 0.25 0.25  0.625 0.625 /
\plot -0.25 0.25  -0.625 0.625 /
\plot -0.25 -0.25  -0.625 -0.625 /
\plot 0.25 -0.25  0.625 -0.625 /
\put {$\chi$} [l] at 0.35 0
\put {$\chi$} [r] at -0.35 0
\put {$t$} [c] at 0 0.40
\put {$t$} [c] at 0 -0.40
\put {$H^\dagger_d$} [l] at 0.65 0.5
\put {$H^c$} [l] at 0.65 -0.5
\put {$H^\dagger_b$} [r] at -0.65 -0.5
\put {$H^a$} [r] at -0.65 0.5
\put {$p_4$} [c] at 0.70 0.70
\put {$p_3$} [c] at 0.70 -0.70
\put {$p_2$} [c] at -0.70 -0.70
\put {$p_1$} [c] at -0.70 0.70
\endpicture$$
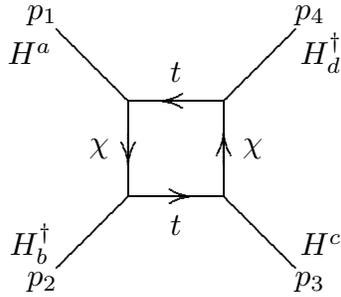
{\bf Figure 5.}  This diagram gives the leading matching contribution to
the operator that violates custodial $SU(2)$ in the effective theory for
energies below the $\chi$ mass. 
\vskip0.125truein

$$\beginpicture
\setcoordinatesystem units <1.00truein,1.00truein>
\setplotarea x from -0.625 to 0.625, y from -0.625 to 0.625
\putrule from 0.375 -0.375 to 0.375 0.375 
\arrow <7pt> [0.2,0.67] from 0.375 -0.08 to 0.375 0.08
\arrow <7pt> [0.2,0.67] from 0.2441 0.2441 to 0.1309 0.1309
\arrow <7pt> [0.2,0.67] from 0.1309 -0.1309 to 0.2441 -0.2441
\plot -0.625 -0.625  0.625 0.625 /
\plot 0.625 -0.625  -0.625 0.625 /
\put {$t_R$} [l] at 0.45 0
\put {$\bullet$} [c] at 0 0
\put {$\psi^3_L$} [c] at 0.125 0.35
\put {$\psi^3_L$} [c] at 0.125 -0.35
\put {$H^\dagger_d$} [l] at 0.55 0.425
\put {$H^c$} [l] at 0.55 -0.425
\put {$H^\dagger_b$} [r] at -0.55 -0.425
\put {$H^a$} [r] at -0.55 0.425
\put {$p_4$} [c] at 0.70 0.70
\put {$p_3$} [c] at 0.70 -0.70
\put {$p_2$} [c] at -0.70 -0.70
\put {$p_1$} [c] at -0.70 0.70
\endpicture
\qquad+\qquad
\beginpicture
\setcoordinatesystem units <1.00truein,1.00truein>
\setplotarea x from -0.625 to 0.625, y from -0.625 to 0.625
\putrule from -0.375 -0.375 to -0.375 0.375 
\arrow <7pt> [0.2,0.67] from -0.375 -0.08 to -0.375 0.08
\arrow <7pt> [0.2,0.67] from -0.2441 0.2441 to -0.1309 0.1309
\arrow <7pt> [0.2,0.67] from -0.1309 -0.1309 to -0.2441 -0.2441
\plot -0.625 -0.625  0.625 0.625 /
\plot 0.625 -0.625  -0.625 0.625 /
\put {$t_R$} [r] at -0.45 0
\put {$\bullet$} [c] at 0 0
\put {$\psi^3_L$} [c] at -0.125 0.35
\put {$\psi^3_L$} [c] at -0.125 -0.35
\put {$H^\dagger_d$} [l] at 0.55 0.425
\put {$H^c$} [l] at 0.55 -0.425
\put {$H^\dagger_b$} [r] at -0.55 -0.425
\put {$H^a$} [r] at -0.55 0.425
\put {$p_4$} [c] at 0.70 0.70
\put {$p_3$} [c] at 0.70 -0.70
\put {$p_2$} [c] at -0.70 -0.70
\put {$p_1$} [c] at -0.70 0.70
\endpicture$$
{\bf Figure 6.}  These graphs gives the leading running contribution to
the operator in the low energy theory that violates custodial $SU(2)$.
\vskip0.125truein

$$\beginpicture
\setcoordinatesystem units <2.00truein,0.20truein>
\setplotarea x from -0.25 to 2.5, y from -3 to 20
\putrule from 0 0 to 2.5 0
\putrule from 0 0 to 0 20
\putrule from 0 0 to -0.05 0
\putrule from 0 2 to -0.05 2
\putrule from 0 4 to -0.05 4
\putrule from 0 6 to -0.05 6
\putrule from 0 8 to -0.05 8
\putrule from 0 10 to -0.05 10
\putrule from 0 12 to -0.05 12
\putrule from 0 14 to -0.05 14
\putrule from 0 16 to -0.05 16
\putrule from 0 18 to -0.05 18
\putrule from 0 20 to -0.05 20
\put {$0$} [r] at -0.1 0
\put {$2$} [r] at -0.1 2
\put {$4$} [r] at -0.1 4
\put {$6$} [r] at -0.1 6
\put {$8$} [r] at -0.1 8
\put {$10$} [r] at -0.1 10
\put {$12$} [r] at -0.1 12
\put {$14$} [r] at -0.1 14
\put {$16$} [r] at -0.1 16
\put {$18$} [r] at -0.1 18
\put {$20$} [r] at -0.1 20
\putrule from 0.5 0 to 0.5 -0.5
\putrule from 1.0 0 to 1.0 -0.5
\putrule from 1.5 0 to 1.5 -0.5
\putrule from 2.0 0 to 2.0 -0.5
\putrule from 2.5 0 to 2.5 -0.5
\put {$0.5$} [c] at 0.5 -1
\put {$1.0$} [c] at 1.0 -1
\put {$1.5$} [c] at 1.5 -1
\put {$2.0$} [c] at 2.0 -1
\put {$2.5$} [c] at 2.5 -1
\put {$m_\chi$} [c] at -0.35 10.5
\put {(TeV)} [c] at -0.35 9.5
\put {$m_{\rm higgs}$ (TeV)} [c] at 1.25 -2.5
\plot 
 0.1226 20.00     0.14   14.34     0.16   11.81     0.18   10.40
 0.20    9.48     0.22    8.82     0.24    8.32     0.26    7.93
 0.28    7.61     0.30    7.34     0.32    7.12     0.34    6.92
 0.36    6.75     0.38    6.60     0.40    6.47     0.42    6.35
 0.44    6.24     0.46    6.14     0.48    6.05     0.50    5.97
 0.52    5.89     0.54    5.82     0.56    5.75     0.58    5.69
 0.60    5.63     0.62    5.58     0.64    5.53     0.66    5.48
 0.68    5.44     0.70    5.39     0.72    5.35     0.74    5.31
 0.76    5.28     0.78    5.24     0.80    5.21     0.82    5.18
 0.84    5.15     0.86    5.12     0.88    5.09     0.90    5.06
 0.92    5.04     0.94    5.01     0.96    4.99     0.98    4.96
 1.00    4.94     1.02    4.92     1.04    4.90     1.06    4.88
 1.08    4.86     1.10    4.84     1.12    4.82     1.14    4.81
 1.16    4.79     1.18    4.77     1.20    4.76     1.22    4.74
 1.24    4.73     1.26    4.71     1.28    4.70     1.30    4.68
 1.32    4.67     1.34    4.66     1.36    4.64     1.38    4.63
 1.40    4.62     1.42    4.61     1.44    4.60     1.46    4.59
 1.48    4.57     1.50    4.56     1.52    4.55     1.54    4.54
 1.56    4.54     1.58    4.53     1.60    4.52     1.62    4.51
 1.64    4.50     1.66    4.49     1.68    4.48     1.70    4.47
 1.72    4.47     1.74    4.46     1.76    4.45     1.78    4.45
 1.80    4.44     1.82    4.43     1.84    4.43     1.86    4.42
 1.88    4.42     1.90    4.41     1.92    4.41     1.94    4.40
 1.96    4.40     1.98    4.40     2.00    4.39     2.02    4.39
 2.04    4.39     2.06    4.39     2.08    4.40     2.08    4.44
 2.06    4.47     2.04    4.49     2.02    4.51     2.00    4.53
 1.98    4.54     1.96    4.56     1.94    4.58     1.92    4.60
 1.90    4.61     1.88    4.63     1.86    4.65     1.84    4.67
 1.82    4.68     1.80    4.70     1.78    4.72     1.76    4.74
 1.74    4.76     1.72    4.77     1.70    4.79     1.68    4.81
 1.66    4.83     1.64    4.85     1.62    4.87     1.60    4.89
 1.58    4.91     1.56    4.93     1.54    4.95     1.52    4.98
 1.50    5.00     1.48    5.02     1.46    5.05     1.44    5.07
 1.42    5.09     1.40    5.12     1.38    5.15     1.36    5.17
 1.34    5.20     1.32    5.23     1.30    5.26     1.28    5.29
 1.26    5.32     1.24    5.35     1.22    5.38     1.20    5.41
 1.18    5.45     1.16    5.48     1.14    5.52     1.12    5.56
 1.10    5.59     1.08    5.63     1.06    5.68     1.04    5.72
 1.02    5.76     1.00    5.81     0.98    5.86     0.96    5.91
 0.94    5.96     0.92    6.02     0.90    6.08     0.88    6.14
 0.86    6.20     0.84    6.27     0.82    6.34     0.80    6.42
 0.78    6.50     0.76    6.58     0.74    6.67     0.72    6.77
 0.70    6.87     0.68    6.98     0.66    7.10     0.64    7.22
 0.62    7.36     0.60    7.51     0.58    7.67     0.56    7.85
 0.54    8.05     0.52    8.28     0.50    8.53     0.48    8.81
 0.46    9.13     0.44    9.51     0.42    9.96     0.40   10.49
 0.38   11.15     0.36   11.99     0.34   13.09     0.32   14.65
 0.30   17.05     0.2865 20.00
/
\setdashpattern <4pt, 2pt>
\plot 
 0.1377 20.00     0.16   14.40     0.18   12.14     0.20   10.80
 0.22    9.89     0.24    9.24     0.26    8.74     0.28    8.34
 0.30    8.01     0.32    7.74     0.34    7.50     0.36    7.30
 0.38    7.13     0.40    6.98     0.42    6.84     0.44    6.72
 0.46    6.61     0.48    6.51     0.50    6.42     0.52    6.34
 0.54    6.26     0.56    6.19     0.58    6.13     0.60    6.07
 0.62    6.02     0.64    5.97     0.66    5.93     0.68    5.90
 0.70    5.87     0.72    5.86     0.72    6.04     0.70    6.15
 0.68    6.26     0.66    6.38     0.64    6.50     0.62    6.62
 0.60    6.75     0.58    6.89     0.56    7.04     0.54    7.21
 0.52    7.39     0.50    7.59     0.48    7.81     0.46    8.06
 0.44    8.34     0.42    8.67     0.40    9.05     0.38    9.50
 0.36   10.05     0.34   10.73     0.32   11.61     0.30   12.81
 0.28   14.58     0.26   17.50     0.252  20.00
/
\setsolid
\linethickness=0.7pt
\putrule from 1.2 14.4 to 1.2 16.6
\putrule from 2.3 14.4 to 2.3 16.6
\putrule from 1.2 14.4 to 2.3 14.4
\putrule from 1.2 16.6 to 2.3 16.6
\linethickness=1.5pt
\putrule from 1.25 16 to 1.5 16
\put {90\% Confidence Level} [l] at 1.55 16
\setdashpattern <4pt, 2pt>
\putrule from 1.25 15 to 1.5 15
\setsolid
\put {68\% Confidence Level} [l] at 1.55 15
\endpicture$$
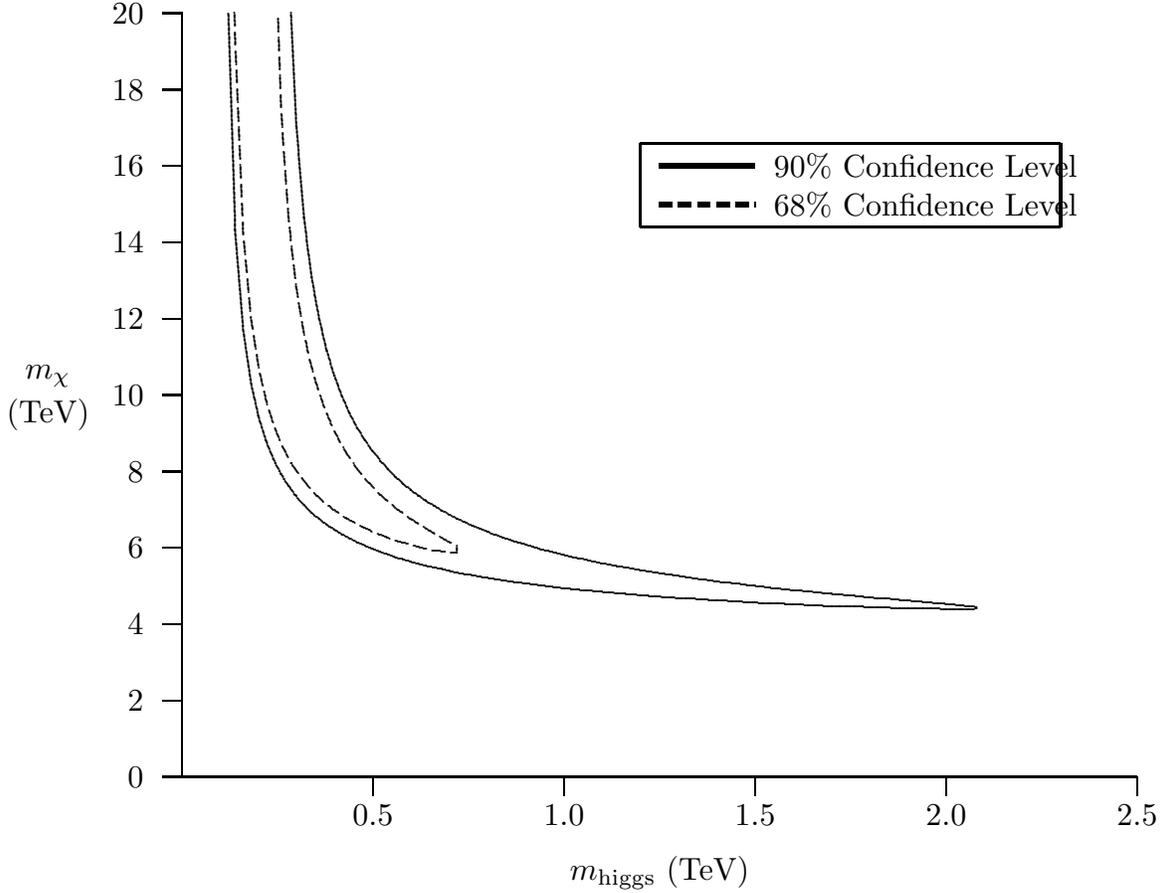
{\bf Figure 7.}  The allowed set of $m_\chi$ and $m_{\rm higgs}$ 
masses based on current precision electroweak tests.
\vskip0.125truein

$$\beginpicture
\setcoordinatesystem units <1.00truein,1.00truein>
\setplotarea x from -1.5 to 1.5, y from 0.0 to 0.5
\putrule from 0.25 0.25 to -0.25 0.25 
\putrule from -0.25 0.25 to -0.25 -0.25 
\putrule from -0.25 -0.25 to 0.25 -0.25 
\putrule from 0.25 -0.25 to 0.25 0.25 
\arrow <7pt> [0.2,0.67] from 0.08 0.25 to -0.08 0.25
\arrow <7pt> [0.2,0.67] from -0.08 -0.25 to 0.08 -0.25
\arrow <7pt> [0.2,0.67] from 0.25 -0.08 to 0.25 0.08
\arrow <7pt> [0.2,0.67] from -0.25 0.08 to -0.25 -0.08
\plot 0.25 0.25  0.625 0.625 /
\plot -0.25 0.25  -0.625 0.625 /
\plot -0.25 -0.25  -0.625 -0.625 /
\plot 0.25 -0.25  0.625 -0.625 /
\put {$\omega$} [l] at 0.35 0
\put {$\chi$} [r] at -0.35 0
\put {$\psi^3_L$} [c] at 0 0.40
\put {$\psi^3_L$} [c] at 0 -0.40
\put {$H^\dagger_{\omega\, d}$} [l] at 0.55 0.375
\put {$H^c_\omega$} [l] at 0.55 -0.375
\put {$H^\dagger_{\chi\, b}$} [r] at -0.55 -0.375
\put {$H^a_\chi$} [r] at -0.55 0.375
\put {$p_4$} [c] at 0.70 0.70
\put {$p_3$} [c] at 0.70 -0.70
\put {$p_2$} [c] at -0.70 -0.70
\put {$p_1$} [c] at -0.70 0.70
\endpicture$$
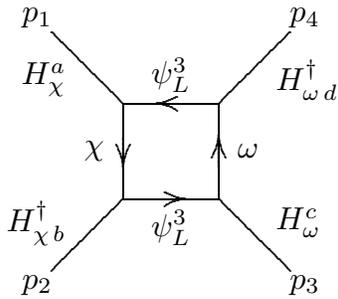
{\bf Figure 8.}  This graph produces the leading matching contribution to
the custodial $SU(2)$-violating operator ${\cal O}_4^{\chi\omega}$.
\vskip0.125truein

$$\beginpicture
\setcoordinatesystem units <1.00truein,1.00truein>
\setplotarea x from -1.5 to 0.75, y from -0.375 to 0.375
\plot 0 -0.375  0 0.375 /
\plot 0.75 0.375  0 0.375  -0.6495 0  0 -0.375  0.75 -0.375 /
\arrow <7pt> [0.2,0.67] from 0 -0.08 to 0 0.08
\arrow <7pt> [0.2,0.67] from 0.295 0.375 to 0.455 0.375
\arrow <7pt> [0.2,0.67] from 0.455 -0.375 to 0.295 -0.375
\hphotondr -0.6495 0 *5 /
\put {$Z$} [r] at -1.50 0
\put {$t$} [l] at 0.10 0
\put {$b$} [l] at 0.85 0.375
\put {$\bar b$} [l] at 0.85 -0.375
\put {$H^\pm$} [r] at -0.32 0.32
\put {$H^\pm$} [r] at -0.32 -0.32
\endpicture
\qquad
\beginpicture
\setcoordinatesystem units <1.00truein,1.00truein>
\setplotarea x from -1.5 to 0.75, y from -0.375 to 0.375
\plot 0 -0.375  0 0.375 /
\plot 0.75 0.375  0 0.375  -0.6495 0  0 -0.375  0.75 -0.375 /
\arrow <7pt> [0.2,0.67] from -0.6495 0 to -0.2555 0.2275
\arrow <7pt> [0.2,0.67] from 0 -0.375 to -0.3940 -0.1475
\arrow <7pt> [0.2,0.67] from 0.295 0.375 to 0.455 0.375
\arrow <7pt> [0.2,0.67] from 0.455 -0.375 to 0.295 -0.375
\hphotondr -0.6495 0 *5 /
\put {$Z$} [r] at -1.50 0
\put {$H^\pm$} [l] at 0.10 0
\put {$b$} [l] at 0.85 0.375
\put {$\bar b$} [l] at 0.85 -0.375
\put {$t$} [r] at -0.32 0.32
\put {$t$} [r] at -0.32 -0.32
\endpicture$$
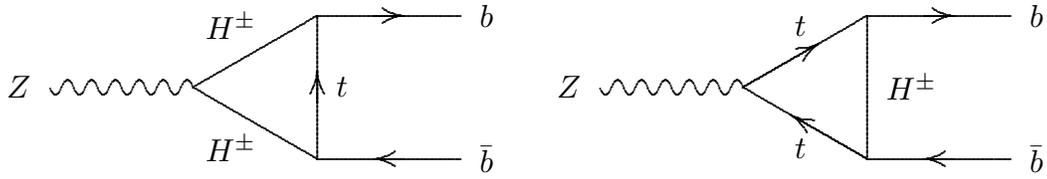
{\bf Figure 9.}  These diagrams produce the dominant one-loop corrections
which involve only the third generation fermions and the appropriate Higgs 
fields.
\vskip0.125truein

$$\beginpicture
\setcoordinatesystem units <1.00truein,1.00truein>
\setplotarea x from -1.5 to 0.75, y from -0.375 to 0.375
\plot 0 -0.375  0 0.375 /
\plot 0.75 0.375  0 0.375  -0.6495 0  0 -0.375  0.75 -0.375 /
\arrow <7pt> [0.2,0.67] from 0 -0.08 to 0 0.08
\arrow <7pt> [0.2,0.67] from 0.295 0.375 to 0.455 0.375
\arrow <7pt> [0.2,0.67] from 0.455 -0.375 to 0.295 -0.375
\hphotondr -0.6495 0 *5 /
\put {$Z$} [r] at -1.50 0
\put {$\chi$} [l] at 0.10 0
\put {$b$} [l] at 0.85 0.375
\put {$\bar b$} [l] at 0.85 -0.375
\put {$H^\pm$} [r] at -0.32 0.32
\put {$H^\pm$} [r] at -0.32 -0.32
\endpicture
\qquad
\beginpicture
\setcoordinatesystem units <1.00truein,1.00truein>
\setplotarea x from -1.5 to 0.75, y from -0.375 to 0.375
\plot 0 -0.375  0 0.375 /
\plot 0.75 0.375  0 0.375  -0.6495 0  0 -0.375  0.75 -0.375 /
\arrow <7pt> [0.2,0.67] from -0.6495 0 to -0.2555 0.2275
\arrow <7pt> [0.2,0.67] from 0 -0.375 to -0.3940 -0.1475
\arrow <7pt> [0.2,0.67] from 0.295 0.375 to 0.455 0.375
\arrow <7pt> [0.2,0.67] from 0.455 -0.375 to 0.295 -0.375
\hphotondr -0.6495 0 *5 /
\put {$Z$} [r] at -1.50 0
\put {$H^\pm$} [l] at 0.10 0
\put {$b$} [l] at 0.85 0.375
\put {$\bar b$} [l] at 0.85 -0.375
\put {$\chi$} [r] at -0.32 0.32
\put {$\chi$} [r] at -0.32 -0.32
\endpicture$$
{\bf Figure 10.}  These diagrams produce the dominant one-loop corrections
which involve the $\chi$ fermion.
\vskip0.125truein

$$\beginpicture
\setcoordinatesystem units <1.00truein,1.00truein>
\setplotarea x from -1.5 to 0.75, y from -0.375 to 0.375
\plot 0 -0.375  0 0.375 /
\plot 0.75 0.375  0 0.375  -0.6495 0  0 -0.375  0.75 -0.375 /
\arrow <7pt> [0.2,0.67] from 0 -0.08 to 0 0.08
\arrow <7pt> [0.2,0.67] from 0.295 0.375 to 0.455 0.375
\arrow <7pt> [0.2,0.67] from 0.455 -0.375 to 0.295 -0.375
\hphotondr -0.6495 0 *5 /
\put {$Z$} [r] at -1.50 0
\put {$\omega$} [l] at 0.10 0
\put {$b$} [l] at 0.85 0.375
\put {$\bar b$} [l] at 0.85 -0.375
\put {$H^0$} [r] at -0.32 0.32
\put {$A^0$} [r] at -0.32 -0.32
\endpicture
\qquad
\beginpicture
\setcoordinatesystem units <1.00truein,1.00truein>
\setplotarea x from -1.5 to 0.75, y from -0.375 to 0.375
\plot 0 -0.375  0 0.375 /
\plot 0.75 0.375  0 0.375  -0.6495 0  0 -0.375  0.75 -0.375 /
\arrow <7pt> [0.2,0.67] from -0.6495 0 to -0.2555 0.2275
\arrow <7pt> [0.2,0.67] from 0 -0.375 to -0.3940 -0.1475
\arrow <7pt> [0.2,0.67] from 0.295 0.375 to 0.455 0.375
\arrow <7pt> [0.2,0.67] from 0.455 -0.375 to 0.295 -0.375
\hphotondr -0.6495 0 *5 /
\put {$Z$} [r] at -1.50 0
\put {$H^0,h^0,A^0$} [l] at 0.10 0
\put {$b$} [l] at 0.85 0.375
\put {$\bar b$} [l] at 0.85 -0.375
\put {$\omega$} [r] at -0.32 0.32
\put {$\omega$} [r] at -0.32 -0.32
\endpicture$$
{\bf Figure 11.}  These diagrams produce the dominant one-loop corrections
which involve the $\omega$ fermion.
\vskip0.125truein

$$\beginpicture
\setcoordinatesystem units <0.06truein,0.60truein>
\setplotarea x from -2.5 to 25, y from -0.8 to 6
\putrule from 0 0 to 25 0
\putrule from 0 0 to 0 6
\putrule from 0 0 to -1.5 0
\putrule from 0 2 to -1.5 2
\putrule from 0 4 to -1.5 4
\putrule from 0 6 to -1.5 6
\put {$0$} [r] at -3 0
\put {$2$} [r] at -3 2
\put {$4$} [r] at -3 4
\put {$6$} [r] at -3 6
\putrule from 5 0 to 5 -0.167
\putrule from 10 0 to 10 -0.167
\putrule from 15 0 to 15 -0.167
\putrule from 20 0 to 20 -0.167
\putrule from 25 0 to 25 -0.167
\put {$5$} [c] at 5 -0.333
\put {$10$} [c] at 10 -0.333
\put {$15$} [c] at 15 -0.333
\put {$20$} [c] at 20 -0.333
\put {$25$} [c] at 25 -0.333
\put {$m_\chi$} [c] at -10.5 3.5
\put {(TeV)} [c] at -10.5 3.167
\put {$m_{h^0}=400\, {\rm GeV}$} [c] at 12.5 0.667
\put {$m_\omega$ (TeV)} [c] at 12.5 -0.833
\plot 
25.00  3.99   24.60  3.99   24.40  3.98   19.00  3.98
18.80  3.99   18.00  3.99   17.80  4.00   17.60  4.00
17.40  4.01   17.20  4.01   17.00  4.02   16.80  4.03
16.60  4.03   16.40  4.05   16.20  4.06   16.00  4.08
15.80  4.10   15.60  4.15   15.60  4.29   15.80  4.36
16.00  4.41   16.20  4.45   16.40  4.49   16.60  4.53
16.80  4.56   17.00  4.59   17.20  4.62   17.40  4.64
17.60  4.67   17.80  4.69   18.00  4.72   18.20  4.74
18.40  4.76   18.60  4.78   18.80  4.80   19.00  4.82
19.20  4.84   19.40  4.86   19.60  4.88   19.80  4.90
20.00  4.91   20.20  4.93   20.40  4.94   20.60  4.96
20.80  4.97   21.00  4.99   21.20  5.00   21.40  5.02
21.60  5.03   21.80  5.04   22.00  5.06   22.20  5.07
22.40  5.08   22.60  5.09   22.80  5.11   23.00  5.12
23.20  5.13   23.40  5.14   23.60  5.15   23.80  5.16
24.00  5.17   24.20  5.18   24.40  5.19   24.60  5.20
24.80  5.21   25.00  5.22   /
\endpicture
\beginpicture
\setcoordinatesystem units <0.06truein,0.60truein>
\setplotarea x from -2.5 to 25, y from -0.8 to 6
\putrule from 0 0 to 25 0
\putrule from 0 0 to 0 6
\putrule from 0 0 to -1.5 0
\putrule from 0 2 to -1.5 2
\putrule from 0 4 to -1.5 4
\putrule from 0 6 to -1.5 6
\put {$0$} [r] at -3 0
\put {$2$} [r] at -3 2
\put {$4$} [r] at -3 4
\put {$6$} [r] at -3 6
\putrule from 5 0 to 5 -0.167
\putrule from 10 0 to 10 -0.167
\putrule from 15 0 to 15 -0.167
\putrule from 20 0 to 20 -0.167
\putrule from 25 0 to 25 -0.167
\put {$5$} [c] at 5 -0.333
\put {$10$} [c] at 10 -0.333
\put {$15$} [c] at 15 -0.333
\put {$20$} [c] at 20 -0.333
\put {$25$} [c] at 25 -0.333
\put {$m_{h^0}=800\, {\rm GeV}$} [c] at 12.5 0.667
\put {$m_\omega$ (TeV)} [c] at 12.5 -0.833
\plot 
25.00  2.98   24.20  2.98   24.00  2.97   22.00  2.97
21.80  2.96   20.00  2.96   19.80  2.95   18.00  2.95
17.80  2.94   13.80  2.94   13.60  2.95   13.40  2.95
13.20  2.96   13.00  2.97   12.80  2.99   12.80  3.07
13.00  3.11   13.20  3.13   13.40  3.16   13.60  3.18
13.80  3.20   14.00  3.21   14.20  3.23   14.40  3.25
14.60  3.26   14.80  3.27   15.00  3.29   15.20  3.30
15.40  3.31   15.60  3.32   15.80  3.33   16.00  3.34
16.20  3.35   16.40  3.36   16.60  3.37   16.80  3.38
17.00  3.39   17.20  3.40   17.40  3.41   17.60  3.41
17.80  3.42   18.00  3.43   18.20  3.44   18.40  3.44
18.60  3.45   18.80  3.46   19.00  3.46   19.20  3.47
19.40  3.47   19.60  3.48   19.80  3.49   20.00  3.49
20.20  3.50   20.40  3.50   20.60  3.51   20.80  3.51
21.00  3.52   21.20  3.52   21.40  3.53   21.60  3.53
21.80  3.54   22.00  3.54   22.20  3.54   22.40  3.55
22.60  3.55   22.80  3.56   23.20  3.56   23.40  3.57
23.80  3.57   24.00  3.58   24.40  3.58   24.60  3.59
25.00  3.59   /
\endpicture
\beginpicture
\setcoordinatesystem units <0.06truein,0.60truein>
\setplotarea x from -2.5 to 25, y from -0.8 to 6
\putrule from 0 0 to 25 0
\putrule from 0 0 to 0 6
\putrule from 0 0 to -1.5 0
\putrule from 0 2 to -1.5 2
\putrule from 0 4 to -1.5 4
\putrule from 0 6 to -1.5 6
\put {$0$} [r] at -3 0
\put {$2$} [r] at -3 2
\put {$4$} [r] at -3 4
\put {$6$} [r] at -3 6
\putrule from 5 0 to 5 -0.167
\putrule from 10 0 to 10 -0.167
\putrule from 15 0 to 15 -0.167
\putrule from 20 0 to 20 -0.167
\putrule from 25 0 to 25 -0.167
\put {$5$} [c] at 5 -0.333
\put {$10$} [c] at 10 -0.333
\put {$15$} [c] at 15 -0.333
\put {$20$} [c] at 20 -0.333
\put {$25$} [c] at 25 -0.333
\put {$m_{h^0}=1200\, {\rm GeV}$} [c] at 12.5 0.667
\put {$m_\omega$ (TeV)} [c] at 12.5 -0.833
\plot 
25.00  2.67   22.80  2.67   22.60  2.66   20.20  2.66
20.00  2.65   17.80  2.65   17.60  2.64   14.20  2.64
14.00  2.65   13.80  2.65   13.60  2.66   13.40  2.66
13.40  2.73   13.60  2.76   13.80  2.77   14.00  2.79
14.20  2.80   14.40  2.81   14.60  2.82   14.80  2.83
15.00  2.84   15.20  2.85   15.40  2.86   15.60  2.87
15.80  2.88   16.00  2.88   16.20  2.89   16.40  2.90
16.60  2.90   16.80  2.91   17.00  2.92   17.20  2.92
17.40  2.93   17.60  2.93   17.80  2.94   18.00  2.95
18.20  2.95   18.40  2.95   18.60  2.96   18.80  2.96
19.00  2.97   19.20  2.97   19.40  2.98   19.60  2.98
19.80  2.99   20.00  2.99   20.20  2.99   20.40  3.00
20.80  3.00   21.00  3.01   21.40  3.01   21.60  3.02
22.00  3.02   22.20  3.03   22.80  3.03   23.00  3.04
23.60  3.04   23.80  3.05   24.40  3.05   24.60  3.06
25.00  3.06   /
\setsolid
\linethickness=0.7pt
\putrule from  2 5.3 to  2 5.7
\putrule from 23 5.3 to 23 5.7
\putrule from  2 5.3 to 23 5.3
\putrule from  2 5.7 to 23 5.7
\linethickness=1.5pt
\putrule from 3.5 5.5 to 11 5.5
\put {90\% CL} [l] at 12.5 5.5
\setdashpattern <4pt, 2pt>
\endpicture$$
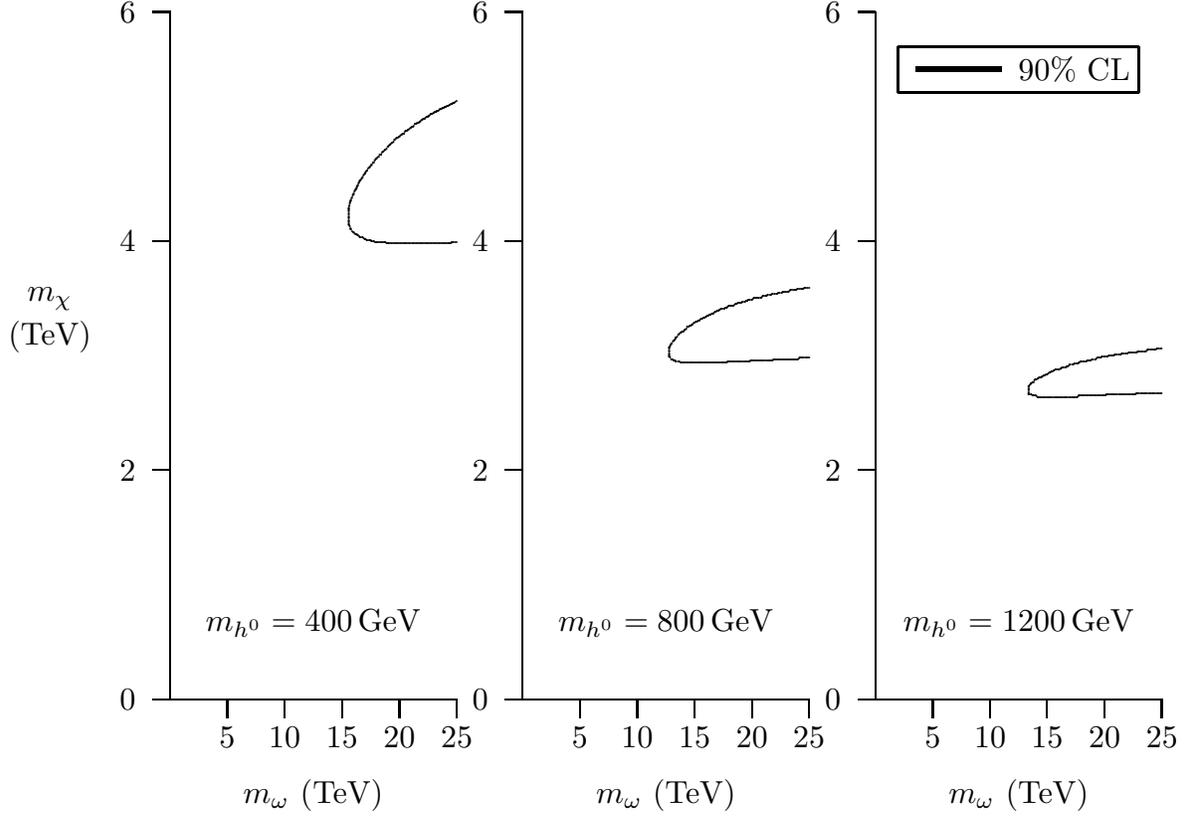
{\bf Figure 12.}  This figure shows the regions in the $\chi$--$\omega$
mass plane allowed by the current precision electroweak experiments when
the massive Higgs fields, $h^0$, $H^0$ and $H^\pm$, have a common mass of
$400$, $800$ and $1200\, {\rm GeV}$ respectively.  In each of these plots,
the light pseudoscalar Higgs field, $A^0$, was given a mass of $100\, {\rm
GeV}$.
\vskip0.125truein

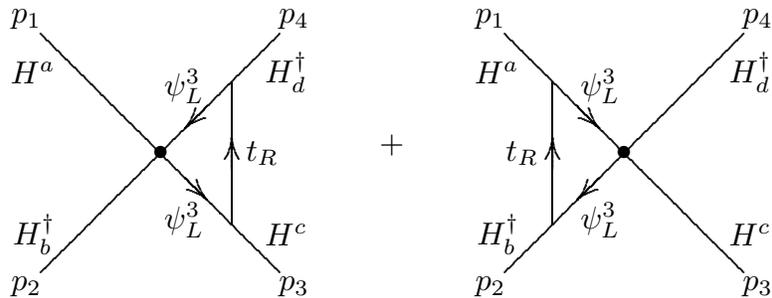
\bye